\documentclass[journal]{IEEEtran}

\usepackage{ragged2e}
\usepackage{graphicx}
\usepackage{float}
\usepackage{subcaption}
\usepackage{amsthm}
\usepackage{color}
\usepackage{tabulary}
\usepackage{verbatim}
\usepackage{enumerate}
\usepackage{rotating}
\usepackage[mathscr]{eucal}
\usepackage{ifthen}
\usepackage{makeidx}
\usepackage{multirow}
\usepackage{amsmath} 
\usepackage{amssymb}
\usepackage{booktabs}
\usepackage{textcmds}
\usepackage{nomencl}
\usepackage{verbatim}
\usepackage{rotating} 
\usepackage[mathscr]{eucal}
\usepackage{ifthen}
\usepackage{makeidx}
\usepackage{multirow}
\usepackage{amsmath} 
\usepackage{enumerate}
\usepackage{amsfonts}
\usepackage{amsmath}
\usepackage{hyperref}
\usepackage{algorithmic}
\usepackage{color,soul}
\usepackage[left=0.67 in,top=0.67 in,right=0.67 in,bottom=0.67 in]{geometry}
\usepackage[normalem]{ulem}
\usepackage{epstopdf}
\usepackage{cite}
\usepackage{cases}
\usepackage{nomencl}
 
\usepackage[font=footnotesize]{caption}
\usepackage[font=footnotesize]{subcaption}
\renewcommand\IEEEkeywordsname{Index Terms}

\renewcommand{\figurename}{Fig.}

\usepackage{algorithm}
\usepackage[table]{xcolor}
\usepackage{algorithmic}
\usepackage{color}

\usepackage{hyperref}
\hypersetup{
    colorlinks=true,
    linkcolor=blue,
    filecolor=magenta,      
    urlcolor=cyan,
}
 
\usepackage{pdfpages}
\pdfoutput=1
\usepackage{fancybox,framed}
\usepackage{csquotes}

\begin{document}
\begin{titlepage}
\centering
\doublebox{%
\begin{minipage}{6in}
\begin{center}
This is the accepted version
\end{center}

\textbf{Live link to published version in IEEE Xplore:} \url{https://ieeexplore.ieee.org/document/8310944}
\newline
\newline
\textbf{Citation to the original IEEE publication:} A. Thakallapelli, S. J. Hossain, and S. Kamalasadan, \enquote{Coherency and Online Signal Selection Based Wide Area Control of Wind Integrated Power Grid}, IEEE Transactions on Industrial Applications, Volume: 54, No: 4, pp. 3712 {-} 3722, Mar. 2018.
\newline
\newline
\textbf{Digital Object Identifier (DOI): 10.1109/TIA.2018.2814561}
\newline
\newline
The following copyright notice is displayed here as per Operations Manual Section 8.1.9 on Electronic Information Dissemination (known familiarly as "author posting policy"):
\newline
\newline
\textcopyright{ 2018 IEEE}. Personal use of this material is permitted. Permission from IEEE must be obtained for all other uses, in any current or future media, including reprinting/republishing this material for advertising or promotional purposes, creating new collective works, for resale or redistribution to servers or lists, or reuse of any copyrighted component of this work in other works.
\end{minipage}}

\end{titlepage}

%
\title{Coherency and Online Signal Selection Based Wide Area Control of Wind Integrated Power Grid}
%
%
%

\author{Abilash ~Thakallapelli,~\IEEEmembership{Student member,~IEEE,}
        Sheikh ~Jakir Hossain,~\IEEEmembership{Student member,~IEEE,}
        and~Sukumar ~Kamalasadan,~\IEEEmembership{Senior Member,~IEEE}
\thanks{A. Thakallapelli, S. J. Hossain, and S. Kamalasadan (corresponding author), are with the Power, Energy and Intelligent Systems Laboratory, Energy Production Infrastructure Center (EPIC) and Department of Electrical Engineering, University of North Carolina at Charlotte, Charlotte, NC 28223 USA (e-mail: athakall@uncc.edu, shossai1@uncc.edu, skamalas@uncc.edu).}
}

\maketitle

\vspace{-37mm}
\begin{abstract}

This paper introduces a novel method of designing Wide Area Control (WAC) based on a discrete Linear Quadratic Regulator (LQR) and Kalman filtering based state-estimation that can be applied for real-time damping of inter-area oscillations of wind integrated power grid.  The main advantages of the proposed method are that the architecture provides: a) online coherency grouping that properly characterizes real-time changes in the power grid, and b) online wide-area signal selection based on residue method for proper selection of the WAC signals.  The proposed architecture can thus accurately monitors changes in the power grid and select the appropriate control signal for more effectively damping the inter-area oscillation when compared to the conventional local signal based Power System Stabilizers (PSS) or offline based WAC designs.  The architecture is tested on a wind integrated two-area system and IEEE  39 bus system in order to show the capability of proposed method.
\end{abstract}


%
\IEEEpeerreviewmaketitle

\section{Introduction}

\IEEEPARstart{I}{nterconnected} power systems exhibit dominant inter-area oscillations with the frequency of oscillation between 0.1-0.8 Hz \cite{ref101}. During inter-area oscillations, coherent generators tend to swing together in groups against other groups based on changes in system conditions \cite{chow2014power}. Inter-area oscillations poses significant problems in the operation of power system as it limits power transfer capability of tie-lines and also deteriorates power system security \cite{ref102}. To suppress the inter-area oscillations, power system stabilizers (PSS) have been designed. PSS provides supplementary damping through synchronous generator exciters \cite{ref103}, high-voltage direct current (HVDC) links \cite{ref104}, and flexible alternating current transmission systems (FACTS) devices \cite{ref105}.

Conventional damping controllers are designed considering specific operating conditions that are typical for the power system \cite{ref2new}. Effectiveness of such controllers decreases when actual operating conditions of the power system deviate from the specified conditions used for the design of such damping controllers. With the increase in integration of renewable energy to the grid, the operating conditions of power grid changes more frequently especially due to the variability in the power output of these renewable resources. Under such variable operating conditions, local robust damping controllers fails to perform well. For example, in \cite{ref106,ref107,ref108} a robust damping controller is designed considering a dominant operating condition with bounded uncertainty to make the controller effective to use under varying operating conditions. As the design is considering a dominant frequency at one operating condition, performance of these controllers may not be optimal as the operating condition deviates from the dominant one. 

Due to the recent developments in wide-area measurement system (WAMS), adaptive damping controller design has drawn a lot of attention. Work performed in \cite{ref109},\cite{ref110} and development of algorithms for applying system identification techniques in power system \cite{ref111} are critical in these aspect.  However, most of the research till now is focused on  tuning of local adaptive controllers while the issue of coordination among all the controller is not addressed well yet. It is worth noting that both adaptive design of local individual controller and the coordination between different controllers can be achieved simultaneously through a properly designed WAC.

There are two major challenges in designing an effective real-time WAC. First, there should be a robust method for obtaining a fast online model of the power system that can capture all low-frequency oscillation modes, and second, there should be an online  method for selecting most observable and controllable signals for the control loop. For successfully managing these challenges, an online coherency grouping method is required, as such grouping reduces the number of signals to be monitored for the design of WAC \cite{ref112}. The reduction in the number of signals decreases computation burden and at the same time facilitates online design and implementation of the WAC. Two types of coherency grouping methods are reported in the literature. The first method is a modal based approach in which system eigenvalues and eigenvectors are analyzed \cite{chow2014power}. Modal based methods are further classified into time domain based slow coherency, and weak link methods. These methods are groups the generators at a specific operating point independent of fault locations. The second method is measurement based \cite{ref2}, which generally uses dynamic response captured from the measured data that characterizes underlying phenomena of oscillations. In this approach coherency grouping is performed by processing measured data using mathematical functions like Fast Fourier transform \cite{reffft}, Hilbert-Huang transform \cite{refhht}, energy function based \cite{refef}, artificial neural networks \cite{refann}, self-organizing maps \cite{refsom1}, graph theory \cite{refgt}, principle component analysis (PCA) \cite{refsom1,refpca}, and k-means clustering \cite{refkme}. Since these methods are based on measurements, it automatically considers and adapts to  changes in system operating conditions. However excessive computational burden and bandwidth requirements limit the success of existing coherency grouping based on measured data. In general, WAC designs are based on the selection of control loops.   In \cite{ref4,ref5} many techniques for selection of control loop for WAC design are illustrated. State-of-the-art methods include residue, and relative gain array (RGA) based approaches. However, they are offline algorithms evaluated based on linearizing the system at a specific operating point and analyzing the eigenvalues and eigenvectors at that instant, making the approach to fail when the operating point changes.

Significant advances have been made on applications of system identification on power system \cite{ref112,ref113,ref114,ref115,ref116}. References \cite{ref113,ref114} shows the application of system identification for estimating electromechanical modes and mode shapes. Subspace state model of the system has also been designed for different operating conditions of the power system by using subspace state-space system identification (N4SID) method \cite{ref116}, or recursive adaptive stochastic subspace identification (RASSI) method \cite{ref115}. However, obtaining subspace state-space model is computationally challenging. Reference \cite{ref112} shows that identifying the power system using auto-regressive moving average exogenous (ARMAX) model structure based  multiple input multiple outputs (MIMO) can accurately capture dynamics of the actual power system. Further in \cite{ref112} an adaptive and coordinated damping controller is designed based on the ARMAX-MIMO model.

In this paper, a novel method of designing online WAC is introduced. This paper is an extended version of \cite{Conf_paper}. In this extended work, three major novel contributions are proposed. First, an online computationally robust coherency grouping of machines is performed based on spectral clustering which uses measurement signals for grouping.  This method considers current system operating condition. In order to overcome the computational burden and to reduce the execution time, only the slow eigenvalues of the Laplacian matrix are used for clustering. Second, the control/loop for WAC is selected based on residue analysis of gain matrix formulated online using MIMO identification which considers the changes in operating conditions of the system. Third, a novel controller is designed based on discrete LQR and state estimation using Kalman filtering. The efficacy of the proposed method is verified by evaluating performance of the architecture to successfully damp the inter-area oscillations on a WTG integrated two-area and IEEE 39 bus system models in real-time using RSCAD/RTDS real-time digital simulator.  To summarize, the major contributions of this work are:
\begin{description}
  \item[$\bullet$] A new coherency grouping method based on online spectral clustering.
  \item[$\bullet$] A new method for online selection of wide-area control loop using MIMO identification.
  \item[$\bullet$] A novel method for WAC  based on discrete LQR and state estimation using Kalman filtering.
\end{description}

The rest of the paper is organized as follows: In section II the proposed framework for the online WAC design is discussed. In section III, the results on a power grid using two-area test system and IEEE 39-bus system are discussed and Section IV concludes the paper.

\section{Proposed Framework for the Online Wide Area Controller}

The proposed framework for the online WAC design involves following steps: a) coherency grouping algorithm, b) signal selection algorithm for WAC based on residue approach and, c) WAC design algorithm.

\subsection{Proposed coherency grouping algorithm}
In this approach coherency grouping of generators is performed based on spectral clustering using measured data points which is the speed of the generators. Data points $x_{1},x_{2},... x_{n}$  for a window length of $n$ are considered for clustering. Further, using the data points a similarity matrix $S\subseteq {R}^{mn}$ is formulated, where $S_{ij}$ gives the relation between $x_{i}$ and $x_{j}$. The similarity matrix information is used to group  $x_{1},x_{2},... x_{n}$  into $k$ clusters. The similarity matrix is a based on a Gaussian function represented as in \eqref{eqn1}

\begin{eqnarray}
S_{ij}=e^{(-\frac{\left\| x_{i}-x_{j}\right\|}{2 \sigma^2})}
\label{eqn1}
\end{eqnarray}
where $\sigma$ is a scaling factor. Here  $S$ is dense and is of the order $n\times n$. The size of $S$ increases with increase in the number of data points under consideration, but this slows the simulation speed. To increase the online clustering speed Nystrom method is used which uses sub-matrix of the dense matrix $\textbf{S}$. Let a sub-matrix $\textbf{A}$ of the dense matrix $\textbf{S}$ can be represented as an $l \times l$ matrix (where $l << n$), sub-matrix $\textbf{B}$  as $l \times (n-l)$ matrix, and $C$  as  $(n-l) \times (n-l)$ matrix. Upon rearranging the columns and rows $\textbf{S}$ can be represented as

\begin{eqnarray}
   \textbf{S}=
   \left[{\begin{array}{cc}
   \textbf{A} & \textbf{B} \\
   \bf{B^T} & \textbf{C} \\
  \end{array} }\right], 
     \textbf{W}=
  \left [{\begin{array}{c}
   \textbf{A}  \\
   \bf{B^T}\\
  \end{array} }\right]
\label{eqn2}
\end{eqnarray}
where $ \textbf{A} \subseteq R^{l \times l}$, $\textbf{B} \subseteq R^{l \times (n-l)}$  , and $\textbf{C} \subseteq R^{(n-l) \times (n-l)}$.   
$S$ is approximated by Nystrom method using  $A$  and $B$. The approximation $S$ is given in \eqref{eqn3}

\begin{align}
\textbf{S} \approx \bar {\textbf{S}} = \left[{\begin{array}{cc}
   \textbf{A} & \textbf{B} \\
   \bf{B^T} & \bf{B^TA^{-1}B} \\
  \end{array} }\right]
\label{eqn3}
\end{align}

The normalized laplacian matrix using the approximated similarity matrix  $\textbf{S}$ can then be represented as \eqref{eqn4}

\begin{align}
\bar{\textbf{L}}= \textbf{I} - {\textbf{D}}^{-\frac{1}{2}}\bar {\textbf{S}}{\textbf{D}}^{-\frac{1}{2}}  
\label{eqn4}
\end{align}
where  $\textbf{D} = \sum_{j=1}^{n} \bar {S_{ij}}$ is a diagonal matrix.
The decomposition of $\bar {\textbf{L}}$  gives the eigenvalues and corresponding eigenvectors according to \eqref{eqn5}. 

\begin{eqnarray}
\bar{\textbf{L}}=  \bar{\textbf{V}} \bar{\bf {\Sigma}} {\bar{\textbf{V}}}^{T}  
\label{eqn5}
\end{eqnarray}
where $\bar {\bf{\Sigma}} $  contains eigenvalues, and $\bar {\bf{V}}$ are the corresponding eigenvectors. If $j$  slow eigenvalues are considered, then the eigenvectors corresponding to $j$  eigenvalues, written as an $R^{n\times j}$ matrix can be formulated as in \eqref{eqn6}

\begin{eqnarray}
\bar{\bf{V}}=  [\vec{v_{1}},\vec{v_{2}}, ...., \vec{v_{n}}]  
\label{eqn6}
\end{eqnarray}	
where $ \vec{v_{i}} \subseteq R^n , i = 1.....j $ and $j$ is the number of eigenvectors. The normalized eigenvector matrix can then be written as

\begin{align}
{\bf{U_{im}}=  \frac{{\bf{\bar{V}_{im}}}}{\sqrt{\sum_{r=1}^{k} {\bar{V}_{ir}}^2}}}, i =1,...,n\quad \textrm{and} \quad m=1,...,j
\label{eqn7}
\end{align}

The $n$ rows of $\bf{U}$ can easily be clustered into  groups using k-means method (see Algorithm 1). In Algorithm-1 the inner-loop repeatedly assigns each set of entities $u_i$ to the closest cluster center $\mu_j$, and recalculate the new cluster center ($\mu_j$) based on the statistical mean of the data points assigned to it. More details of k-means method can be found in \cite{kmean,ref6}.

\begin{algorithm}
\caption{K-means Algorithm}
\begin{algorithmic} 
\STATE 1) Given $\textbf{U}=\{u_1,u_2,...,u_n\}$ (set of entities to be clustered)
\STATE 2)	Select $k$ initial cluster centers $\mu_{1}$, $\mu_{2}$,..., $\mu_{k}$ $\in$ $\mathbb{R}$.
\REPEAT 
\STATE a) For every $i$, set $c_i=\underset{j}{\mathrm{argmin}} \|{u_{i}}-\mu_{j}\|^2$ 
\STATE b) For every $j$, set $\mu_j = \frac{\sum_{i=1}^{n} 1\{c_i=j\}u_i}{\sum_{i=1}^{n} 1\{c_i=j\}}$
\UNTIL{converged}
\end{algorithmic}
\end{algorithm}

\begin{figure}[!h]
\centering
\includegraphics[width=3.5in]{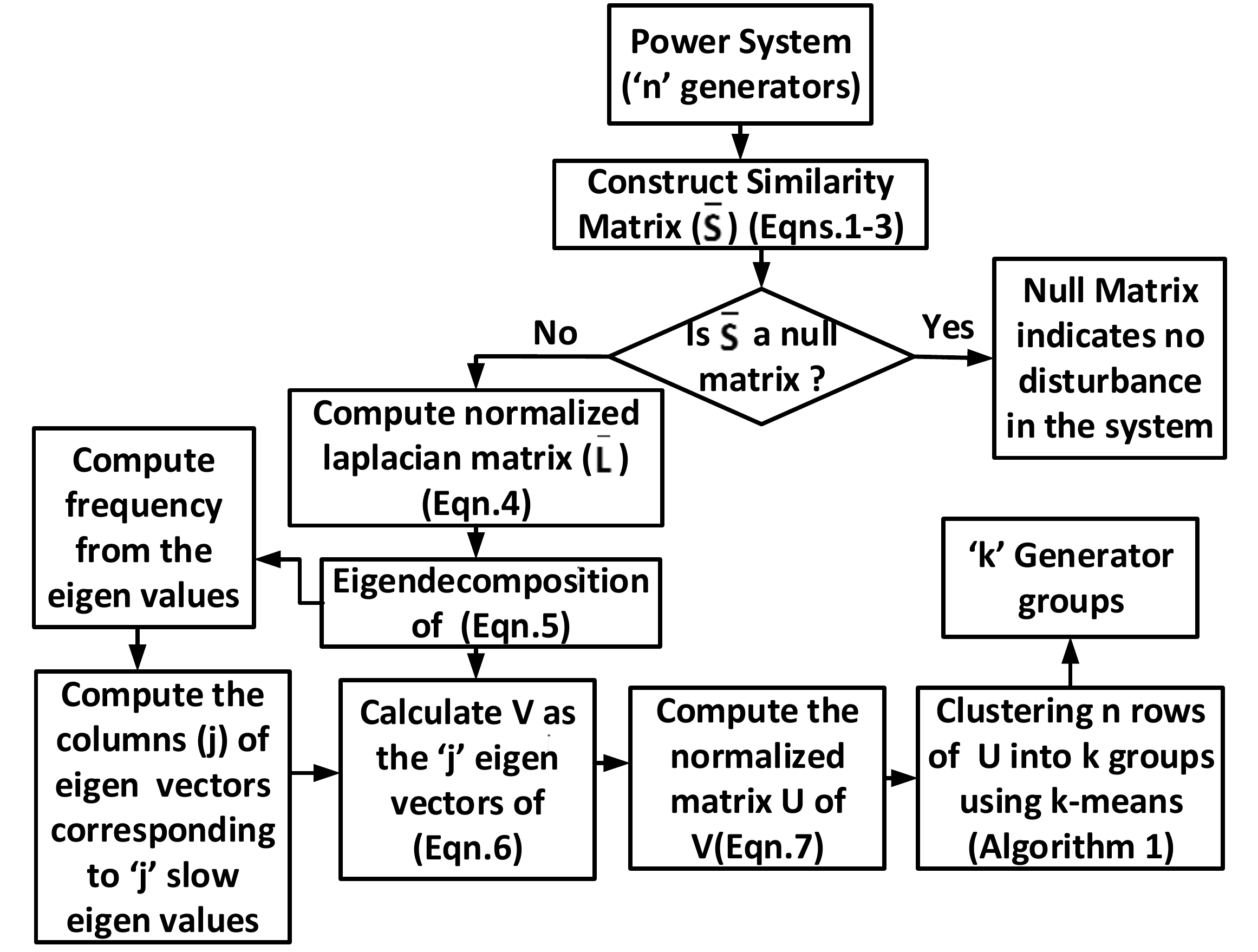}
\caption{Flow chart of the proposed coherency grouping algorithm.}
\label{fig_cluster_flow}
\end{figure}

\begin{figure}[!h]
\centering
\includegraphics[width=0.475\textwidth,height=3in]{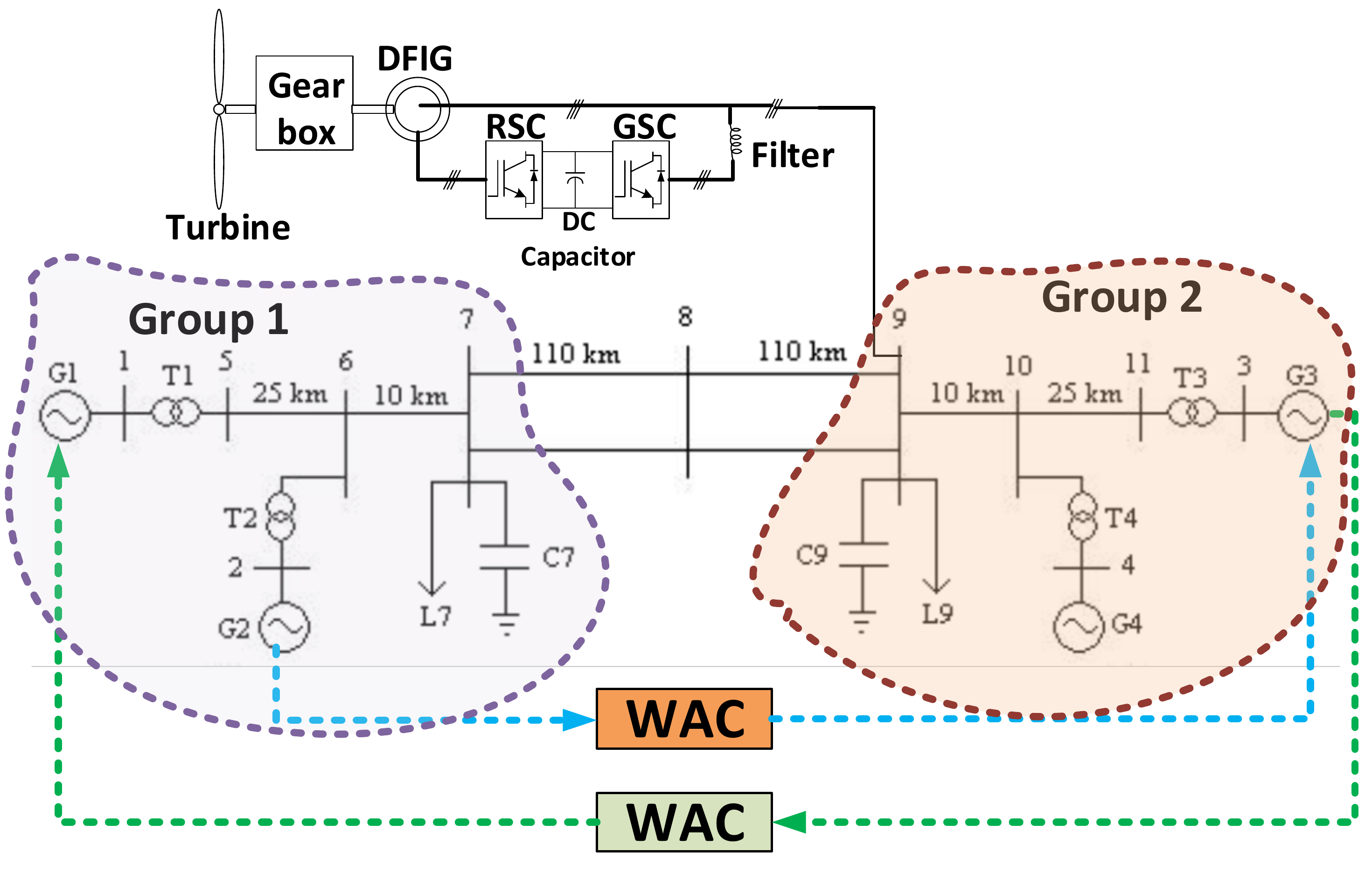}
\caption{Two-area study system model.}
\label{fig_two_area}
\vspace{-2mm}
\end{figure}

\begin{figure}[!h]
\centering
\includegraphics[width=0.475\textwidth]{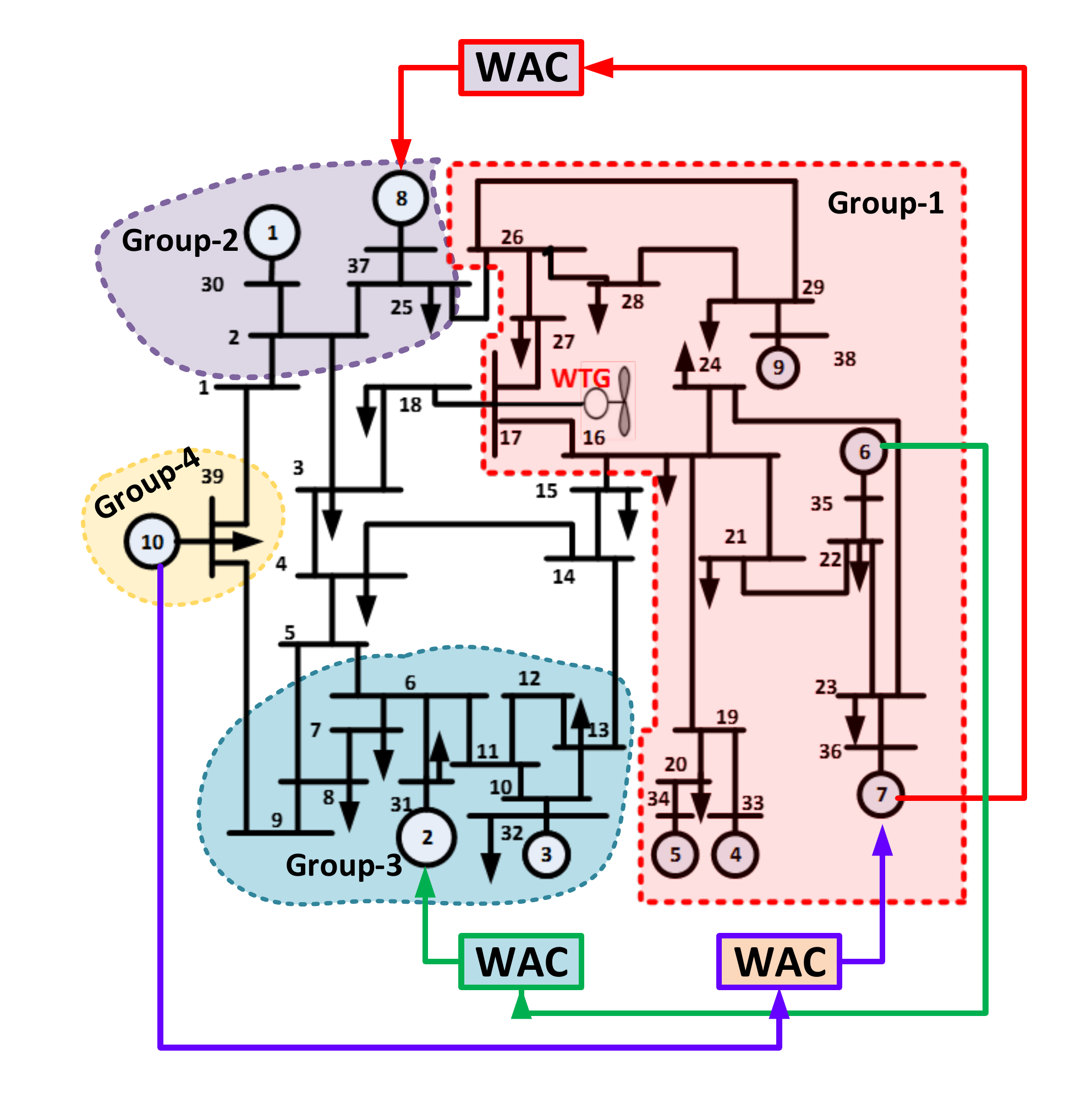}
\caption{IEEE-39 bus test system model (Fault on Bus-14).}
\label{fig_39_bus}
\vspace{-2mm}
\end{figure}

\begin{table}[!h]
\renewcommand{\arraystretch}{1.3}
\centering
\caption{COMPARISON OF COHERENCY GROUPING METHODS}
\label{table_1}
\begin{tabular}{*9c} 
\toprule
 Test System & Slow Coherency & Spectral Clustering\\
\hline
Two Area & Group-1:1,2  & Group-1:1,2\\
     & Group-2:3,4  & Group-2:3,4\\
     & $t$ = 0.4826s & $t$ = 0.021s\\ 
\hline
IEEE 39-BUS & Group-1:4,5,6,7  & Group-1:4,5,6,7,9\\
     & Group-2:1,8,9  & Group-2:1,8\\
     & Group-3:2,3  & Group-3:2,3\\
     & Group-4:10  & Group-4:10\\
     & $t$ = 0.8138s & $t$ = 0.0423s\\
\hline
*$t$ = Computational Time
\end{tabular}
\end{table}

\begin{figure}[h]
\centering
\includegraphics[width=3.5in]{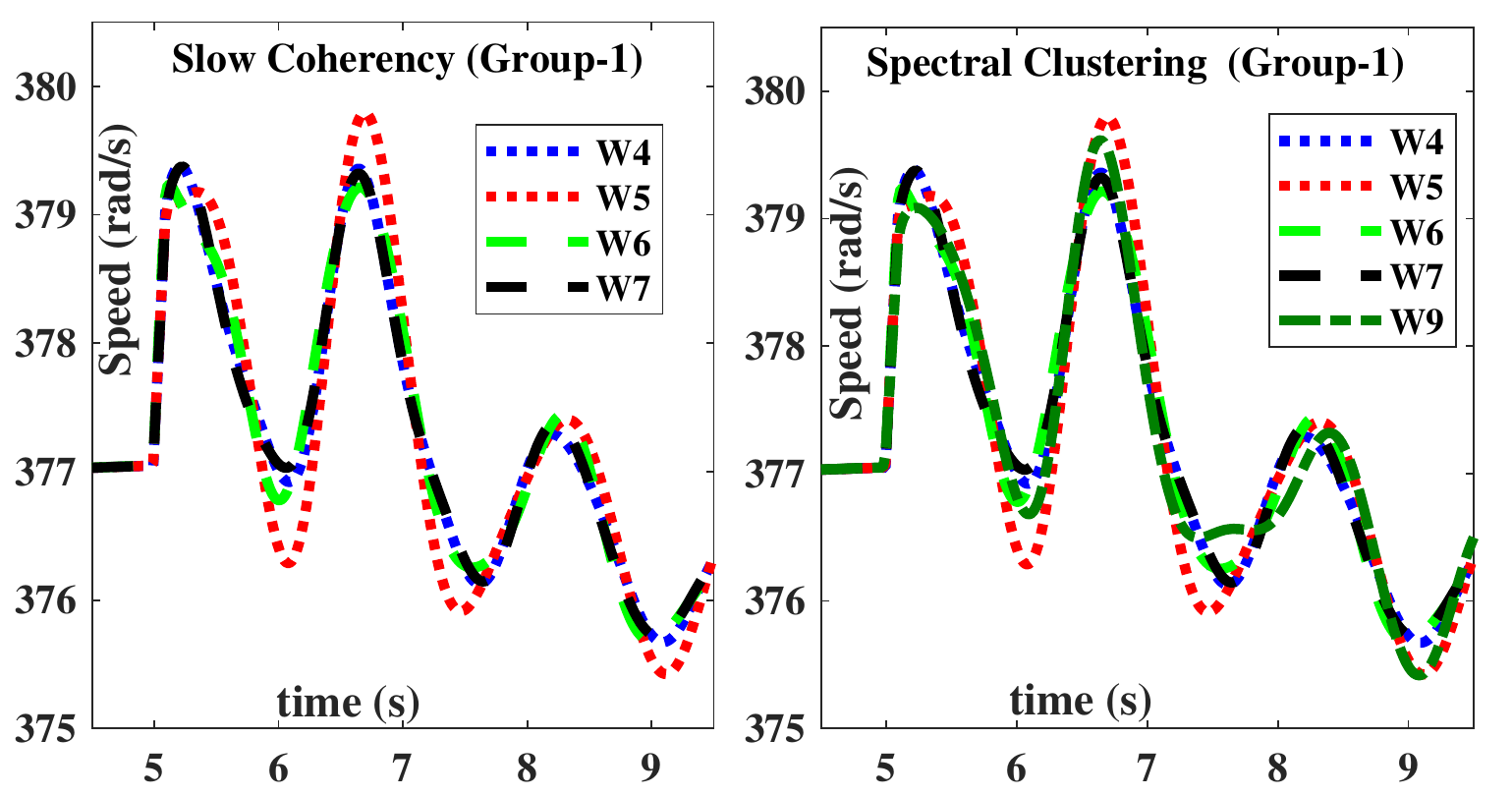}
\caption{Coherency group-1}
\label{fig_4a}
\vspace{-3mm}
\end{figure}

\begin{figure}[h]
\centering
\includegraphics[width=3.5in]{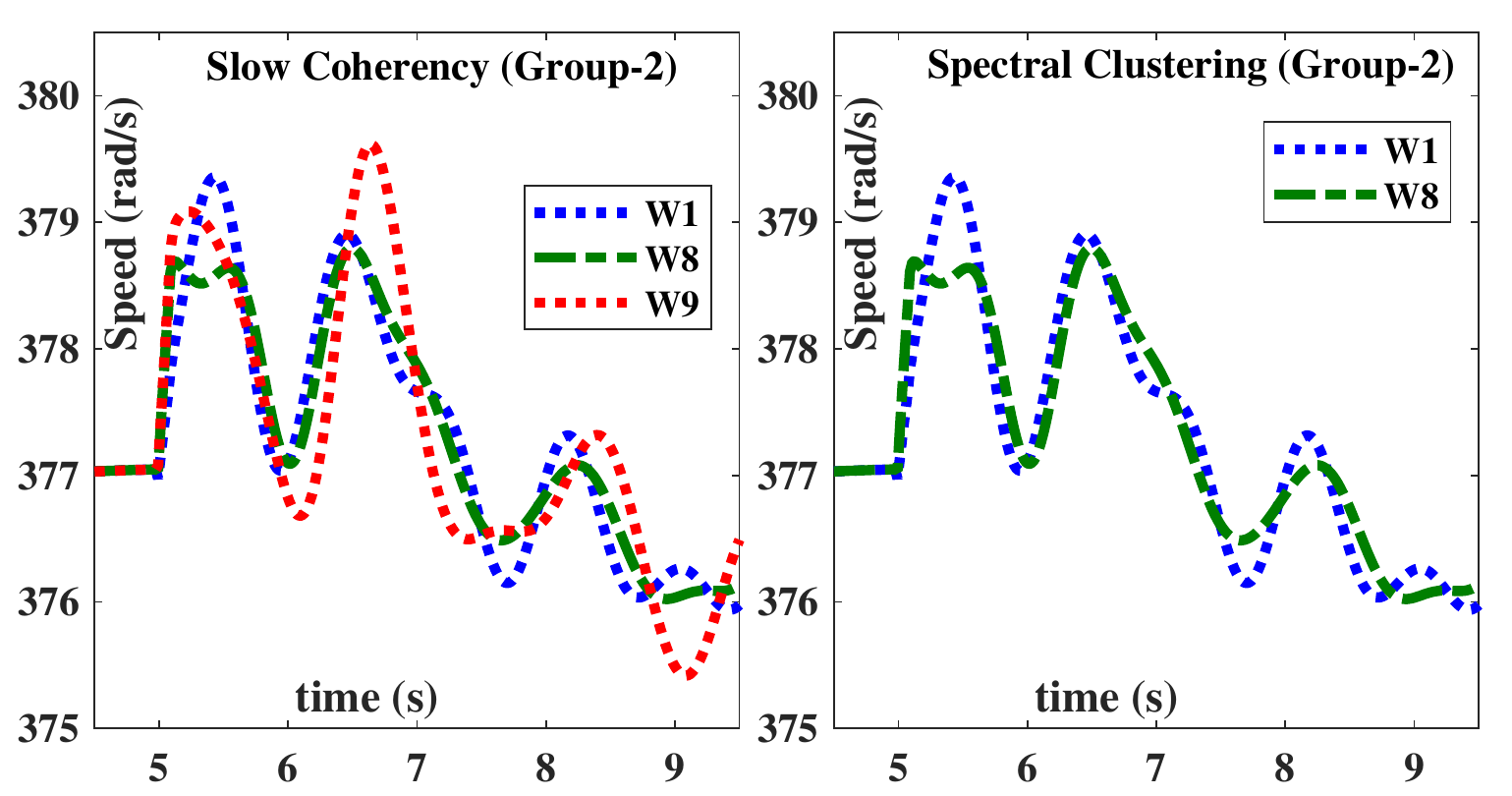}
\caption{Coherency group-2}
\label{fig_5a}
\vspace{-3mm}
\end{figure}

\begin{figure}[h]
\centering
\includegraphics[width=3.5in]{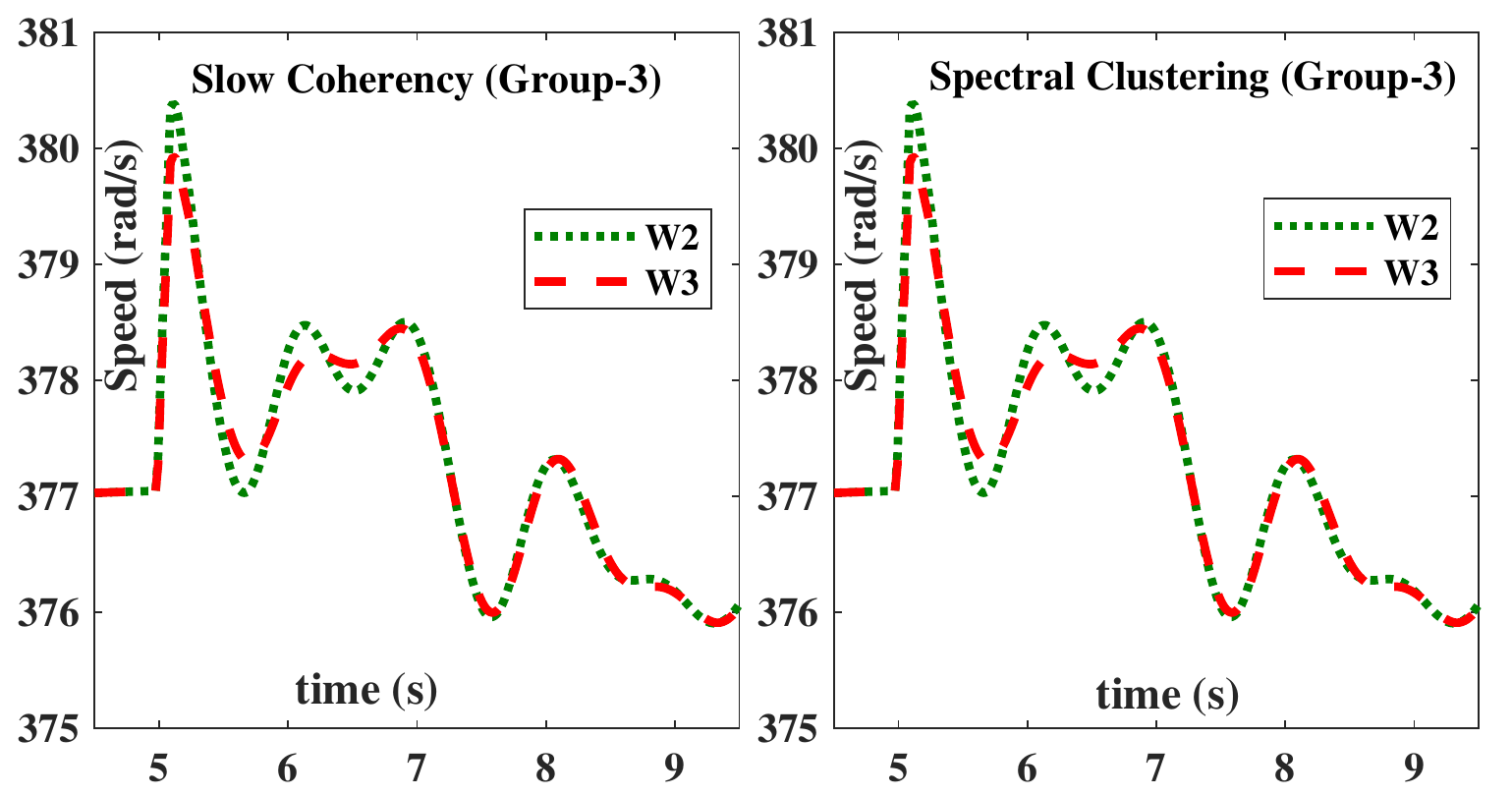}
\caption{Coherency group-3}
\label{fig_6a}
\vspace{-3mm}
\end{figure}

\begin{figure}[h]
\centering
\includegraphics[width=3.5in]{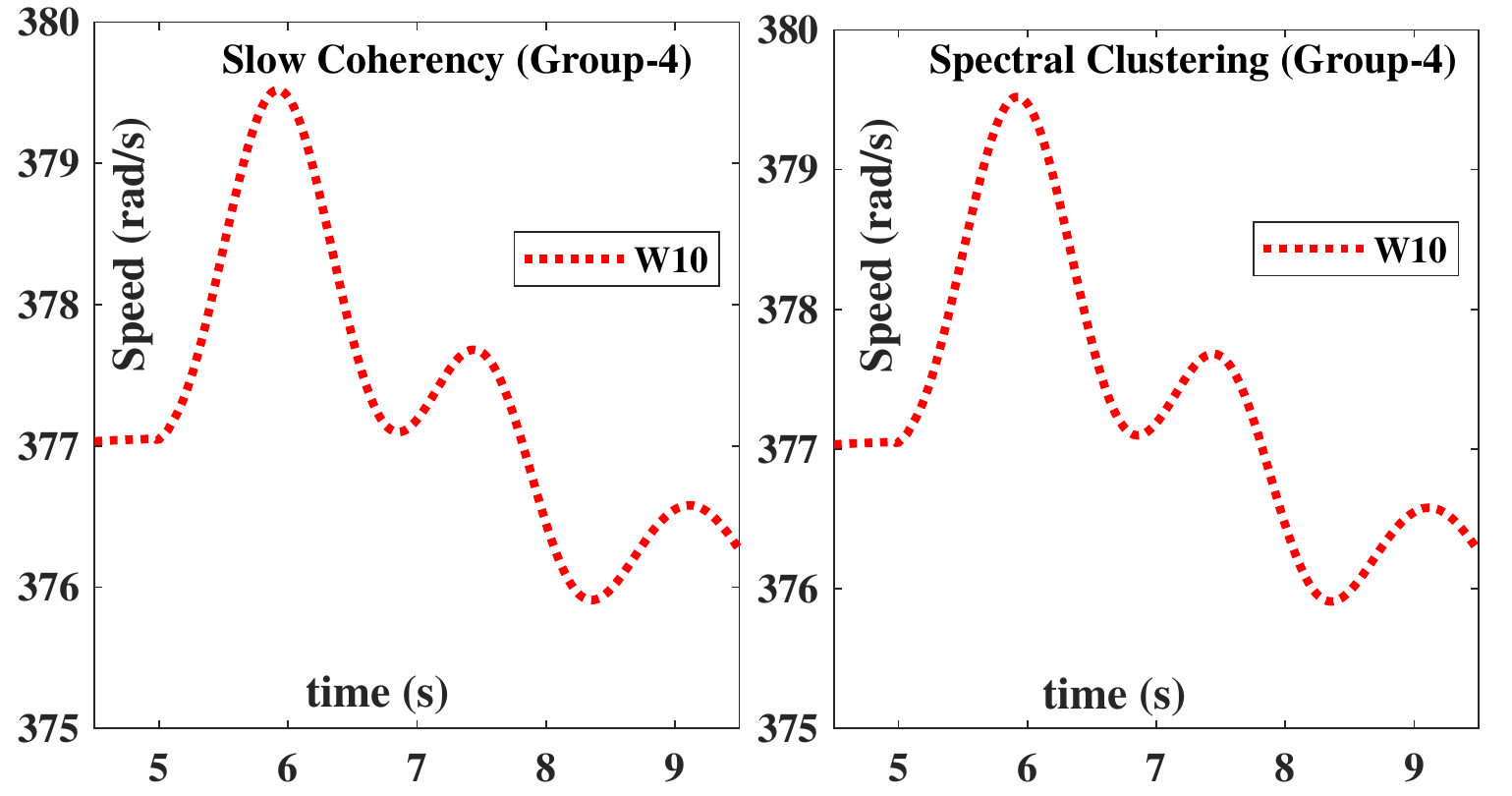}
\caption{Coherency group-4}
\label{fig_7a}
\vspace{-3mm}
\end{figure}

Fig. \ref{fig_cluster_flow} shows the flow chart of the proposed algorithm. The efficacy of the proposed algorithm is verified by implementing the architecture on Kundur two area (Fig. \ref{fig_two_area}) and IEEE-39 bus power system models (Fig. \ref{fig_39_bus}), and then comparing the method with state-of-the-art offline slow coherency based grouping.  For two-area system a 3-ph fault is created on bus-9 for a duration of 0.1 sec. The coherent groups obtained based on the proposed algorithm and comparisons are shown in Table \ref{table_1}. It can be seen that the proposed online method provides same grouping as offline method but with lower computational time. Also, for IEEE 39 bus system, group 1 is different with generator 9 included in group 1. This is wrongly interpreted in the slow coherency method.  Fig. \ref{fig_4a} to Fig. \ref{fig_7a} shows the comparison of coherency grouping for a fault on bus-14.

\begin{figure*}[!h]
\centering
\includegraphics[width=\textwidth]{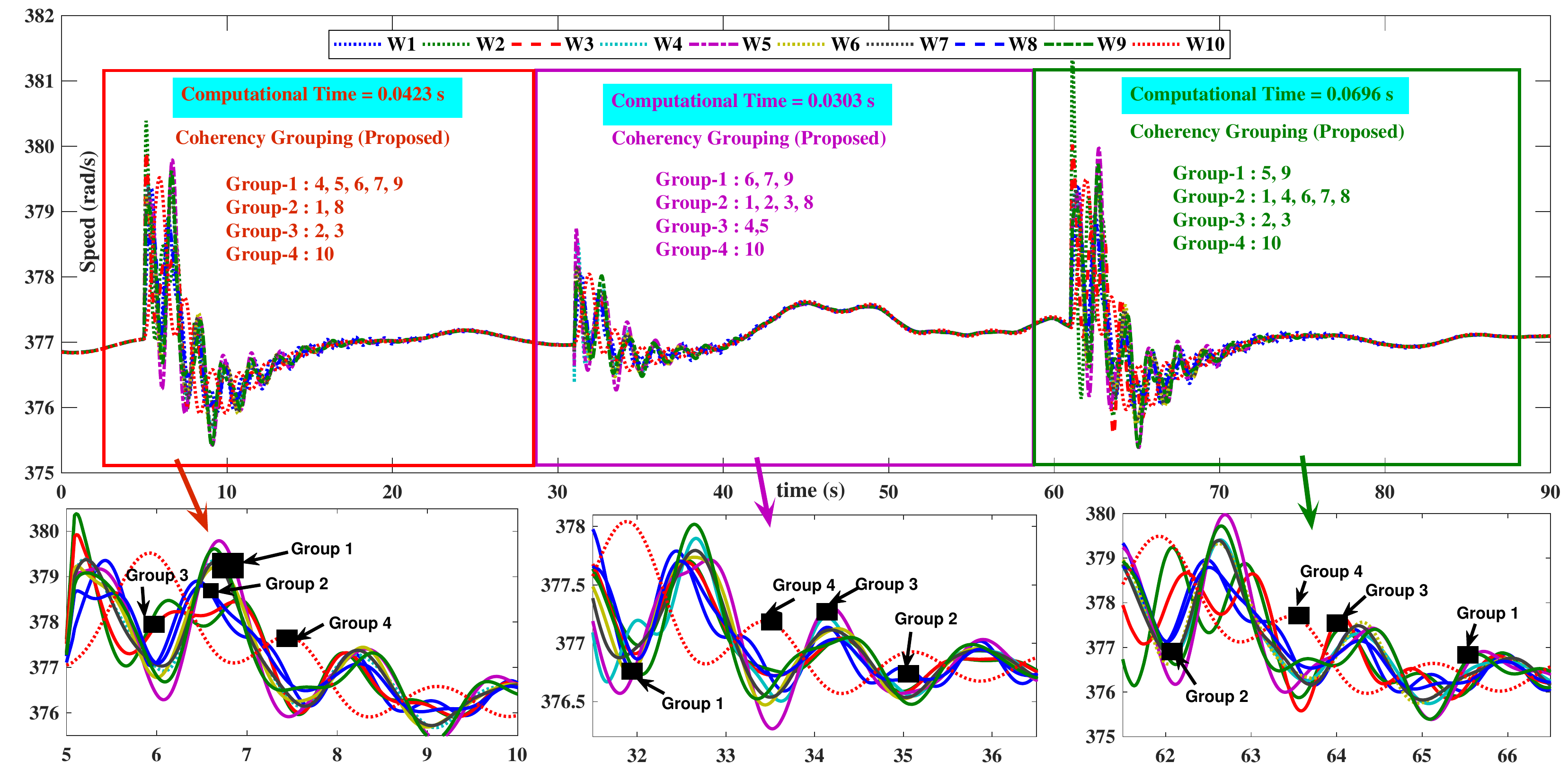}
\caption{Coherency grouping for different operating conditions.}
\label{fig_aa}
\end{figure*}

Further to validate the algorithm for changing operating conditions in real-life, a sequence of disturbances are studied at various locations of IEEE-39 bus system. For this, 3-ph faults are created for a duration of 0.1 sec at bus-14, bus-19, and bus-6 at 5, 31 and 61 sec respectively. Table \ref{table_1a} shows the coherency grouping comparison of 39-bus system for various operating conditions. Fig. \ref{fig_aa} shows the coherency grouping for various operating conditions for IEEE-39 bus system.  It can be observed that with the proposed coherency grouping method the generators can be grouped based on current operating condition whereas in offline methods like slow coherency the generator grouping does not change irrespective of changing operating conditions. The dynamics observed at different operating conditions in Fig. \ref{fig_aa} supports this statement. It is also worth noting that the computational time required for the proposed online coherency grouping method is much less confirming that the algorithm is feasible to implement on a real-life system.

 \begin{figure}[!h]
\centering
\includegraphics[width=3.5in]{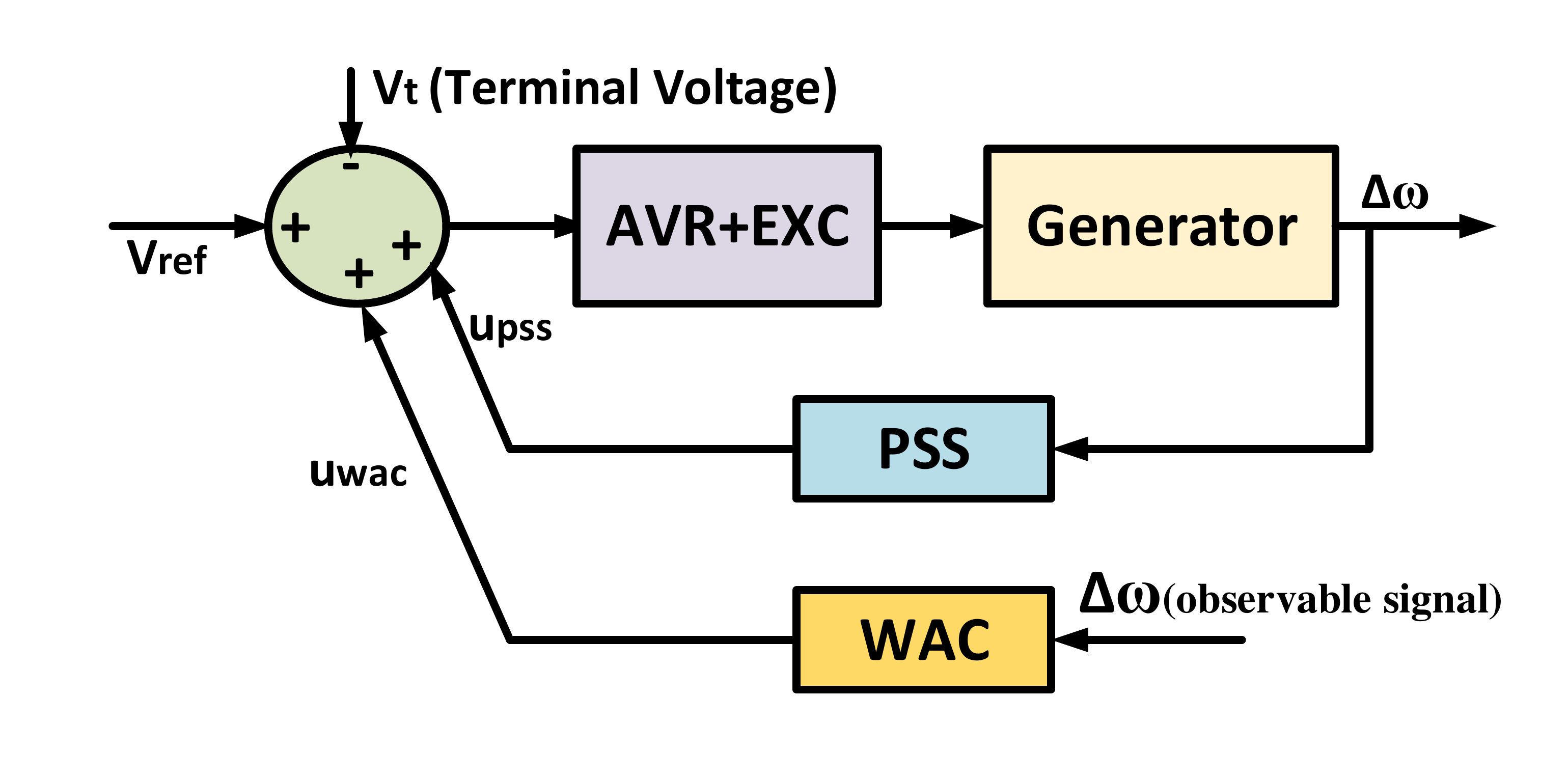}
\caption{Input/Output signal for transfer function estimation.}
\label{fig_8}
\end{figure} 

\begin{table*}[!h]
\renewcommand{\arraystretch}{1.3}
\centering 
\caption{COHERENCY GROUPING OF IEEE-39 BUS SYSTEM  (SPECTRAL CLUSTERING) }
\label{table_1a}
\begin{tabular}{|c|c|c|c|}
\hline
\toprule
 \multicolumn{4}{|c|}{Fault Bus}\\
\midrule
No of Clusters (k)  & Bus-14 & Bus-19 & Bus-6 \\
\hline
4 & Computational Time = 0.0423s & Computational Time = 0.0303s & Computational Time = 0.0696s\\
  & Group-1:4,5,6,7,9 & Group-1:6,7,9 & Group-1:5,9 \\ 
  & Group-2:1,8 & Group-2:1,2,3,8 & Group-2:1,4,6,7,8 \\ 
  & Group-3:2,3 & Group-3:4,5 & Group-3:2,3 \\ 
  & Group-4:10 & Group-4:10 & Group-4:10 \\ 
\hline
\end{tabular}
\end{table*}

For MIMO identification of the power system, input and output signals need to be selected. The output signal is taken as the speed of synchronous machines and the input signal is taken as the voltage input to the automatic voltage regulator (AVR) and exciter of the synchronous machines. The measurement points for input and output is shown in Fig. \ref{fig_8}. If there are $m$ output and $p$ input signals selected, then  \eqref{eqn9} represents  the MIMO system representation of the power system in discrete time domain. For signal selection, the MIMO identification is performed first to identify the transfer functions in \eqref{eqn9}. A Recursive least square identification (RLS) algorithm \cite{ref7,ref8,ref9,ref11,ref12} is proposed for the MIMO system identification.

 Further, based on the MIMO identification, transfer function of the power system can be developed as shown in equation \eqref{eqn10} \cite{ref7,ref8,ref9,ref11,ref12}. In \eqref{eqn10}, $\Delta\omega_{m}$ is the difference between rated and actual speed of the generator,  $u_{p}$  is the input signal (Fig. \ref{fig_8}), $k$ the order of transfer function and  $h$ is the element number in matrix. From \eqref{eqn10}, it can be seen that all transfer functions in \eqref{eqn9} have same denominator coefficients but unique numerator coefficients. This ensures that coupling between different loops are considered while identifying the MIMO system. Also, this indicates that the power systems modes as seen from any input are same however their corresponding residues differ. The identification of MIMO system is  divided into three steps which are discussed in the following subsections.

 \begin{eqnarray}
     \begin{split}
    \bf{\Delta\omega(z)} &=\bf{G}(z)\bf{U}(z)\\
    \end{split}
     \label{eqn9}
    \end{eqnarray}
    where
 \begin{eqnarray}
     \begin{split}
  \bf{\Delta\omega}(z) &= \left[{\begin{array}{c}
   \Delta\omega_{1}(z) \\
   . \\
   . \\
   \Delta\omega_{m}(z) 
  \end{array} }\right],
  \bf{U}(z) = \left[{\begin{array}{c}
   u_{1}(z) \\
   . \\
   . \\
   u_{p}(z) \\
  \end{array} }\right]\\
   \bf{G}(z) &=  \left[{\begin{array}{cccc}
   G_{11}(z) &. &. &G_{1p}(z)\\
   .  &. &. &.\\
   .  &. &.  &.\\
   G_{m1}(z) &. &. &G_{mp}(z)\\
  \end{array} }\right]
 \end{split} \IEEEnonumber
\end{eqnarray}
 
 \begin{eqnarray}
 \begin{split}
  \bf{G_{mp}}(z) &=\frac{\Delta\omega_{m}(z)}{u_{p}(z)} \\
                 &=\frac{b_0^h+b_1^hz^{-1}+...+b_k^hz^{-k}}{1+a_1z^{-1}+a_2z^{-2}+...+ a_kz^{-k}}
 \end{split}
  \label{eqn10}
\end{eqnarray}

\subsection{Signal selection algorithm for WAC design }

\subsubsection{Step-1} 
For $N$  observation window length, and $j$ samples, rewrite  \textbf{\eqref{eqn10}} as shown in \eqref{eqn11}.
 
\begin{eqnarray}
  \bf{X_{His}^h} =\bf{X_{Num}^h} +\bf{X_{Den}^h}
  \label{eqn11}
\end{eqnarray} 	
where
\begin{eqnarray}
 \bf{X_{His}^h} = \left[{\begin{array}{c}
   \Delta\omega_{1}(z) \\
   . \\
   . \\
   \Delta\omega_{m}(z) 
  \end{array} }\right]\IEEEnonumber
\end{eqnarray}

\begin{eqnarray}
  \bf{X_{Num}^h} = \left[{\begin{array}{ccc}
   u_{p}(j-1) &. &u_{p}(j-k)\\
   .  &.  &.\\
   .  &.  &.\\
   u_{p}(j-N) &. &u_{p}(j-N+1-k)\\
  \end{array} }\right]\left[{\begin{array}{c}
   b_0^h \\
   . \\
   . \\
   b_k^h \\
  \end{array} }\right]\IEEEnonumber
\end{eqnarray} 	

\begin{eqnarray}
  \bf{X_{Den}^h} = \left[{\begin{array}{ccc}
   \Delta\omega_{m}(j-1) &. &\Delta\omega_{m}(j-k)\\
   .  &.  &.\\
   .  &.  &.\\
   \Delta\omega_{m}(j-N) &. &\Delta\omega_{m}(j-N+1-k)\\
  \end{array} }\right]\left[{\begin{array}{c}
   a_1 \\
   . \\
   . \\
   a_k \\
  \end{array} }\right]\IEEEnonumber
\end{eqnarray} 	
\begin{table*}[!t]
\renewcommand{\arraystretch}{1.3}
\centering
\caption{CONTROL LOOP BASED ON RESIDUE APPROACH}
\label{table_3}
\begin{tabular}{*9c}
\hline
\toprule
Type & \multicolumn{4}{c}{Residue analysis using MIMO identification}&  \multicolumn{4}{c}{Residue analysis of state space matrices}\\
\midrule
 Coherent groups & \multicolumn{2}{c}{Group 1} & \multicolumn{2}{c}{Group 2} & \multicolumn{2}{c}{Group 1} & \multicolumn{2}{c}{Group 2}\\
\hline
Speed/Field voltage &\cellcolor{yellow}  $u_1$ & $u_2$ & $\cellcolor{green} u_3$ & $u_4$ & $u_1$ & $u_2$ & $u_3$ & $u_4$\\
\hline
$\Delta\omega_1$  & \cellcolor{yellow} 0.5640 & 0.5280 & \cellcolor{green} 0.6398 & 0.6323 & 0.4973 & 0.6081 & 0.5263 & 0.6537\\
\hline
\cellcolor{green} $\Delta\omega_2$  &\cellcolor{yellow}  0.4323 & \cellcolor{green} 0.6571 & \cellcolor{green}  {\color{red}1.0000} & 0.6385 & 0.3551 & 0.4342 & 0.3758 & 0.4668\\
\hline
\cellcolor{yellow} $\Delta\omega_3$  &\cellcolor{yellow}  {\color{red}0.8758} & 0.7143 & 0.7366 & 0.9602 & 0.7608 & 0.9303 & 0.8051 & 1.0000\\
\hline
$\Delta\omega_4$  & 0.7093 & 0.6609 & 0.8584 & 0.7826 & 0.6733 & 0.8232 & 0.7125 & 0.8850\\
\hline
 & \multicolumn{4}{c}{Computational Time (Signal Selection) = 1.28s}&  \multicolumn{4}{c}{Computational Time (Signal Selection) = 0.078s}\\
  & \multicolumn{4}{c}{Computational Time (WAC Output Signal) = 0.0116s} &  \\
\hline
\end{tabular}
\end{table*}
\subsubsection{Step-2} 
Concatenate $\bf{X_{His}^h}$  for all input-output combinations  as shown in \eqref{eqn12}

\begin{eqnarray}
 \left[{\begin{array}{c}
   \bf{X_{His}^1} \\
   \bf{X_{His}^2} \\
   . \\
   \bf{X_{His}^h}\\
  \end{array} }\right] = \left[{\begin{array}{c}
   \bf{X_{Num}^1} \\
   \bf{X_{Num}^2} \\
   . \\
   \bf{X_{Num}^h}\\
  \end{array} }\right]+ \left[{\begin{array}{c}
   \bf{X_{\Delta\omega}^1} \\
   \bf{X_{\Delta\omega}^2} \\
   . \\
   \bf{X_{\Delta\omega}^h}\\
  \end{array} }\right]\left[{\begin{array}{c}
   a_1 \\
   . \\
   . \\
   a_k \\
  \end{array} }\right]
    \label{eqn12}
\end{eqnarray}

\subsubsection{Step-3} 
Calculate denominator coefficients and numerator coefficients iteratively. For this, in the first iteration numerator coefficients are initialized. Then the denominator coefficients $(a_1,a_2... a_k)$ are calculated by applying least squares technique to \eqref{eqn13}. Further the numerator coefficients $(b_0^h,b_1^h... b_k^h)$  are calculated again as shown in \eqref{eqn14}

\begin{eqnarray}
\left[{\begin{array}{c}
   \bf{X_{\Delta\omega}^1} \\
   \bf{X_{\Delta\omega}^2} \\
   . \\
   \bf{X_{\Delta\omega}^h}\\
  \end{array} }\right]\left[{\begin{array}{c}
   a_1 \\
   . \\
   . \\
   a_k \\
  \end{array} }\right] =  \left[{\begin{array}{c}
   \bf{X_{His}^1} \\
   \bf{X_{His}^2} \\
   . \\
   \bf{X_{His}^h}\\
  \end{array} }\right] - \left[{\begin{array}{c}
   \bf{X_{Num}^1} \\
   \bf{X_{Num}^2} \\
   . \\
   \bf{X_{Num}^h}\\
  \end{array} }\right]
    \label{eqn13}
\end{eqnarray}

\begin{eqnarray}
\left[{\begin{array}{c}
  \bf{X_{up}^h}
  \end{array} }\right]\left[{\begin{array}{c}
   b_0^h \\
   . \\
   . \\
   b_k^h \\
  \end{array} }\right] = \left[{\begin{array}{c}
  \bf{X_{His}^h}
  \end{array} }\right] - \left[{\begin{array}{c}
  \bf{X_{\Delta\omega}^h}
  \end{array} }\right]\left[{\begin{array}{c}
   a_1 \\
   . \\
   . \\
   a_k \\
  \end{array} }\right]
    \label{eqn14}
\end{eqnarray}
The numerator and denominator are calculated iteratively until the desired tolerance is achieved such that $||\left([\bf{X_{His}^h}-(\bf{X_{Num}^h}+\bf{X_{Den}^h})])\right|| \leq  0.0001 $.
The MIMO identification flow is shown in Algorithm 2.

\begin{algorithm}
\caption{MIMO Identification-Algorithm}
\begin{algorithmic} 
\STATE 1)	Initialize numerator coefficients $(b_0^h,b_1^h... b_k^h)$ of (9)
\WHILE{$tol \leq 0.0001$}

{\STATE a)	Calculate denominator coefficients $(a_1,a_2... a_k)$ (12) 
\STATE b)	Using denominator coefficients obtained in previous step, calculate numerator coefficients (13)
\STATE c)	tol = $norm([\bf{X_{His}^h}-(\bf{X_{Num}^h}+\bf{X_{Den}^h})])$} \ENDWHILE 
\end{algorithmic}
\end{algorithm}

After identifying the entire system with different inputs and outputs for $G_{mp}(z)$, \eqref{eqn10} is formulated. Then $G_{mp}(z)$ is converted from discrete to continuous time domain  ($G_{mp}(s)$) and transformed into a partial fraction form, such that \eqref{eqn10} takes the form of \eqref{eqn15}.

\begin{align}
\begin{split}
   \bf{G_{mp}}(s) &=\frac{\Delta\omega_{m}(s)}{u_{p}(s)} \\
                  &= \frac{r_{mp}(1)}{s-p_1} + \frac{r_{mp}(2)}{s-p_2} + ..+ \frac{r_{mp}(j)}{s-p_j}+ k_{mp}(s) 
\end{split}
  \label{eqn15}   
\end{align}
where $r_{mp}(j)$ is the residue of $G_{mp}(s)$ corresponding to the pole $p_j$. The residue $r_{mp}(j)$ provides information about how the mode $p_j$ is affected by input $u_p$ and how observable is it from $\Delta\omega_m$. This indicates that residue is a measure of joint controllability and observability index, where larger the value of residue, the stronger is the control loop. 

The validity of the proposed algorithm for signal selection is verified by implementing this approach on a wind integrated two-area power system model. For this, a dominant mode of 0.6038 Hz is observed. Table \ref{table_3} shows the comparison of proposed signal selection algorithm with the residue analysis \cite{ref4} of state space matrices corresponding to the dominant mode. Since, there are two coherent groups in two-area system, two WAC are required; one in each group. Generators 1 and 2 are in one coherent group and the remaining generators are in second coherent group. It is worth noting that, as the value of residue gets larger, the stronger is the control loop to damp oscillations. 

The control loop required to damp the observed mode can be derived from Table \ref{table_3} as follows. In coherent group-1 the residue value is largest between speed of generator-3 ($\Delta\omega_3$) and input signal of generator-1 ($u_1$ ), so generator-1 is the most controllable machine. Likewise, in coherent group-2 the most observable signal is $\Delta\omega_2$, and generator 4 is most controllable machine. Table \ref{table_2a} shows the WAC control loop for the case of IEEE-39 bus system with a fault on bus-14 and with four clusters. In group-1, generator 8 is the most controllable machine and the most observable signal is $\Delta\omega_7$, in group-2, generator 10 is the most controllable machine and the most observable signal is $\Delta\omega_1$, in group-3, generator 2 is the most controllable machine and the most observable signal is $\Delta\omega_6$, and in group-4, generator 7 is the most controllable machine and the most observable signal is $\Delta\omega_{10}$. However the residue of control loop for group-2 is very low, which can be verified from the fact that the inertia of generator-10 is very high compared to other generators and so it is the less controllable machine. Hence the controller for group-2 is ignored.

\subsection{WAC design architecture}

\begin{figure}[h]
\centering
\includegraphics[width=3.5in]{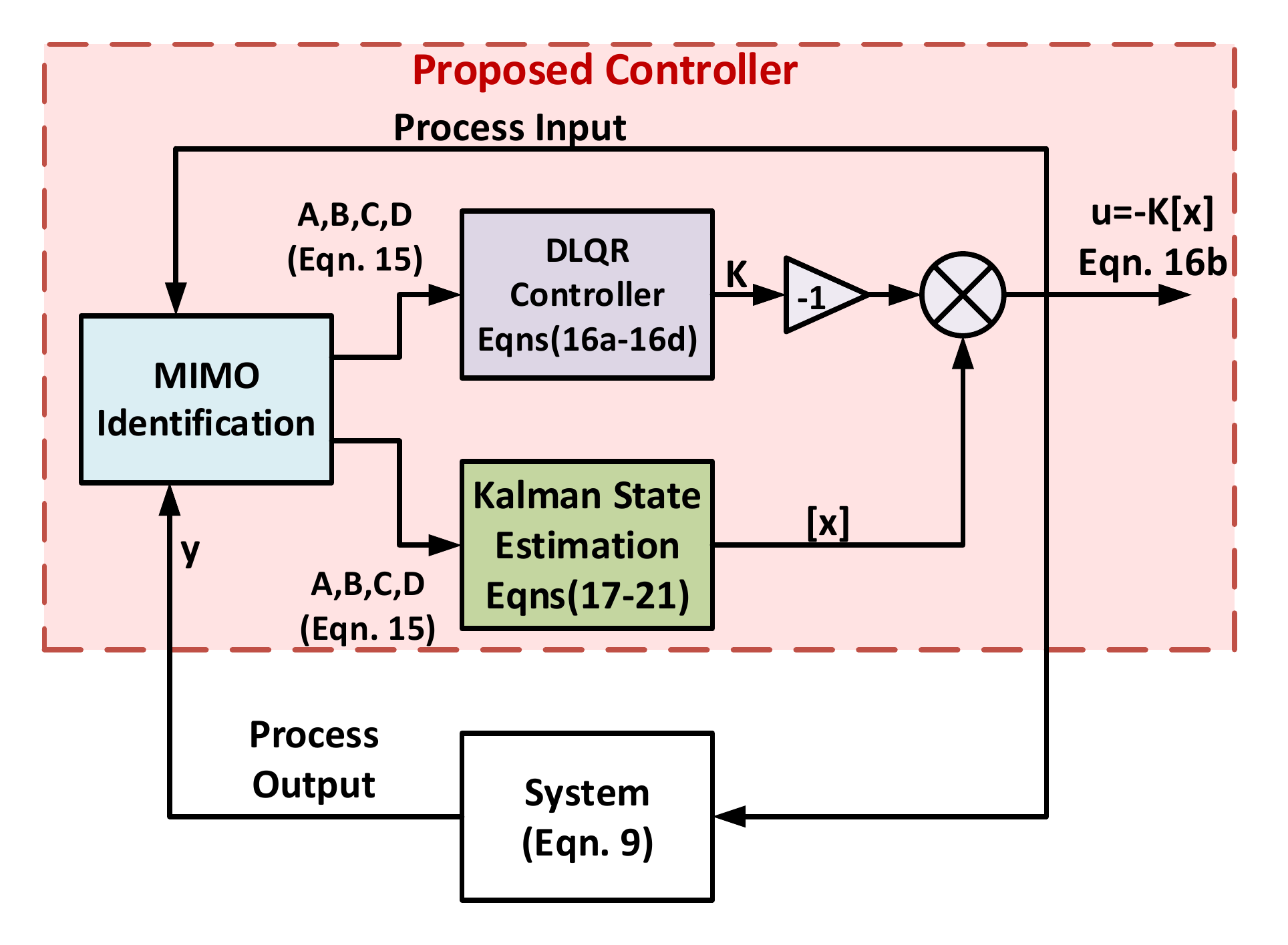}
\caption{Proposed wide area controller.}
\label{fig_9}
\end{figure}

The proposed WAC algorithm has three parts 1) identification of control loop transfer function, 2) Discrete-time Linear Quadratic Regulator (DLQR), and 3) state estimation using Kalman filtering. The details are discussed below. Fig. \ref{fig_9} shows the architecture of the proposed WAC design.

\subsubsection{Identification of control loop transfer function}
In the proposed WAC, the state space matrices of the control loop between process input $(u_p)$ and the process output $\Delta\omega_m$ is calculated using the transfer function coefficients obtained from \eqref{eqn9} to  \eqref{eqn14}. In general, the order $k$ is not known a priori, so $k$ is assumed to be a large number which is limited by number of samples $l$ and computational burden. Then, modes with negligible residues are discarded \cite{ref13new} and new order $p$ is identified (i.e the order of \eqref{eqn10} is reduced from $k$ to $p$). The discrete state space matrices are formulated using \eqref{eqn10} which can be written as in \eqref{eqn16}.

\begin{align}
\begin{split}
   \bf{x}(p+1)&= \bf{Ax}(p)+\bf{Bu}(p) \\
              \bf{y(p)} &= \bf{Cx}(p)+\bf{Du}(p) 
\end{split}
  \label{eqn16}   
\end{align}

where $\bf{A}$ is the state matrix, $\bf{B}$  is the input matrix,  $\bf{C}$ is the output matrix, $\bf{D}$  is the feedforward matrix, $u$ represents state vector, and $y$ represents input and output respectively.

\subsubsection{Discrete-time linear quadratic regulator}
In the proposed algorithm, the discrete-time linear quadratic regulator is used to minimize the cost function for the discrete-time linear system model as described in \eqref{eqn16}. The cost function\cite{ref13} is represented as shown in \eqref{eqn17} where  $N$ is the sample horizon, $\bf{Q} = C^TC$ and $R = \rho I (\rho > 0) $  are the weight matrices. The cost function is minimized by calculating the optimal gain matrix $\bf{K}$, such that the optimal control sequence is given by  \eqref{eqn18}
and the optimal control gain $\bf{K}$ is given by \eqref{eqn19}. $\bf{P}$ is the solution of the discrete time algebraic Riccati equation, which is represented as \eqref{eqn20} by iterating backward in time until the solution converges, initially during start of the iteration $\bf{P_p} = 0$.

\begin{subequations}
\begin{align}
   \bf{J} &= \bf{\sum_{p=0}^{N} ({x_p^T}Qx_p + {u_p^T}Ru_p)}  \label{eqn17}\\  
   u_{p} &= -\bf{K_p} x_p  \label{eqn18}\\
   \bf{K_p} &= \bf{(R + B^TP_{p+1}B)^{-1}B^T P_{p+1} A}  \label{eqn19}\\
   \begin{split}
 \bf{P_{p-1}}&= \bf{Q +A^T P_p A }\\ 
                &\bf{-A^T P_p B(R+B^T P_p B)^{-1} B^T P_p A}        \label{eqn20}
 \end{split}
\end{align}
\end{subequations}

\subsubsection{State estimation using Kalman filtering}
The optimal control sequence as given by \eqref{eqn18} requires states of the system to be estimated, which is done using Kalman filtering technique. The state space representation in  \eqref{eqn16} obtained from RLS identification is used for this estimation.
The predictor step for state vector and co-variance is given by \eqref{eqn21} and \eqref{eqn22}  respectively.
\begin{eqnarray}
   \bar{\bf{x}} = \bf{A}\bf{x} + \bf{B}\bf{u}
  \label{eqn21}\\
     \bar{\bf{L}} = \bf{A}\bf{L}\bf{A^T} + \bf{Q}
  \label{eqn22}  
\end{eqnarray} 
where $L$ is the covariance of state vector estimate and $Q$ is the process noise covariance. The Kalman gain factor ($G$) is calculated as shown in \eqref{eqn23}
 
\begin{eqnarray}
   \bf{G} = \bar{\bf{L}}\bf{H^T} \bf{{(H\bar{L}{H^T}+R)}^{-1}}
  \label{eqn23}  
\end{eqnarray}	
where $H$ is the observation matrix and $R$ is the measurement noise covariance. The corrector step is given by \eqref{eqn24} and \eqref{eqn25}.
 
\begin{eqnarray}
   \bf{x} = \bar{\bf{x}} + \bf{G(z-H\bar{x})} 
  \label{eqn24}\\
     \bf{L} = \bar{\bf{L}}- \bf{KH\bar{L}}
  \label{eqn25}  
\end{eqnarray}	
Additional details of state estimation using Kalman filtering are provided in \cite{ref14}.

\section{Simulation Results }

\begin{figure}[!h]
\centering
\includegraphics[width=3.5in]{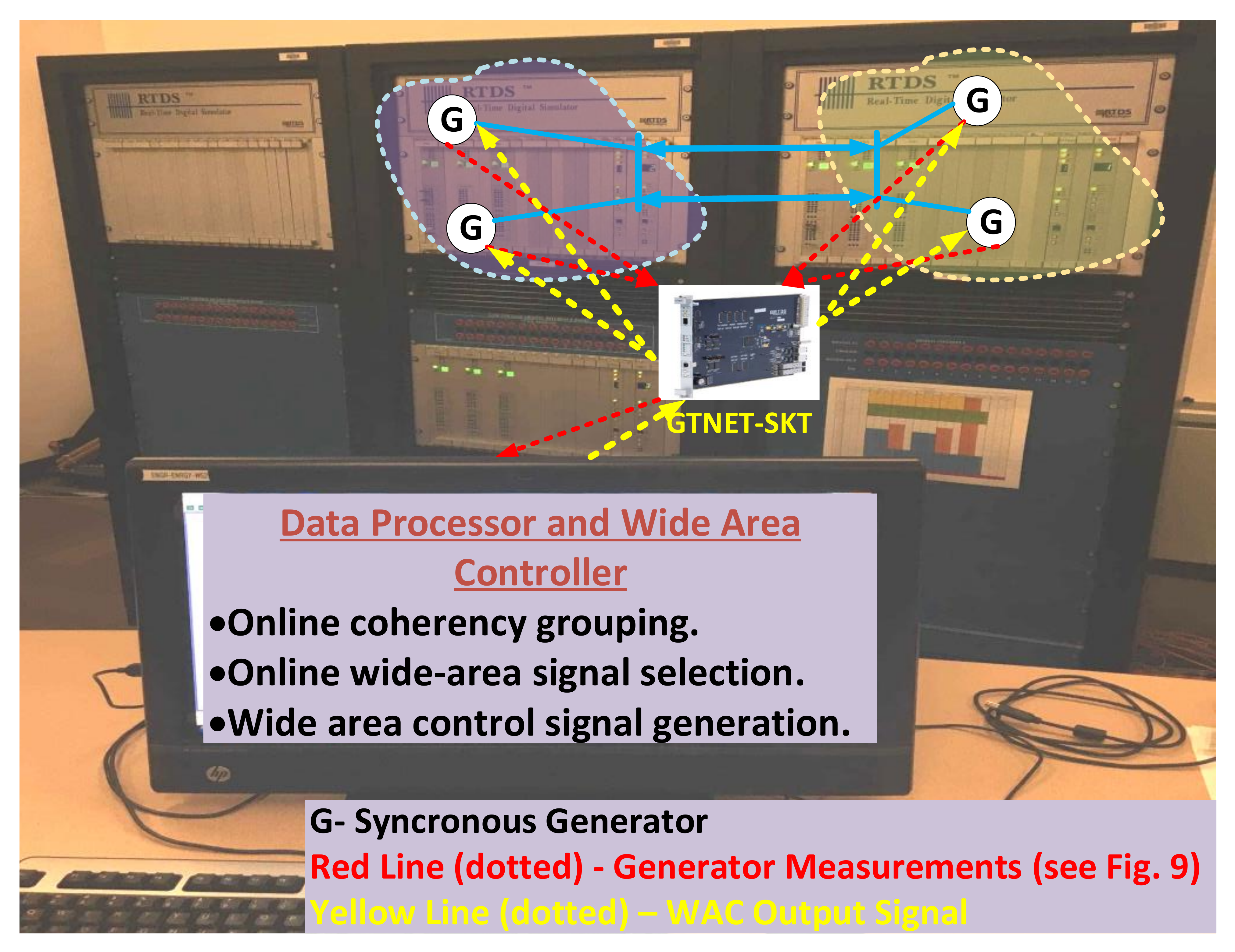}
\caption{Experimental test bed.}
\label{fig_11}
\end{figure}

The proposed controller is implemented on the two-area (Fig. 1) and IEEE 39-bus (Fig. 2) system interconnected with a 150MW WTG \cite{ref15} with each WTG is rated at 2MW. The experimental test-bed is as shown in Fig. \ref{fig_11}. 

 \subsection{Implementation test results using two-area system}
\begin{figure}[!t]
\centering
\includegraphics[width=3.5in]{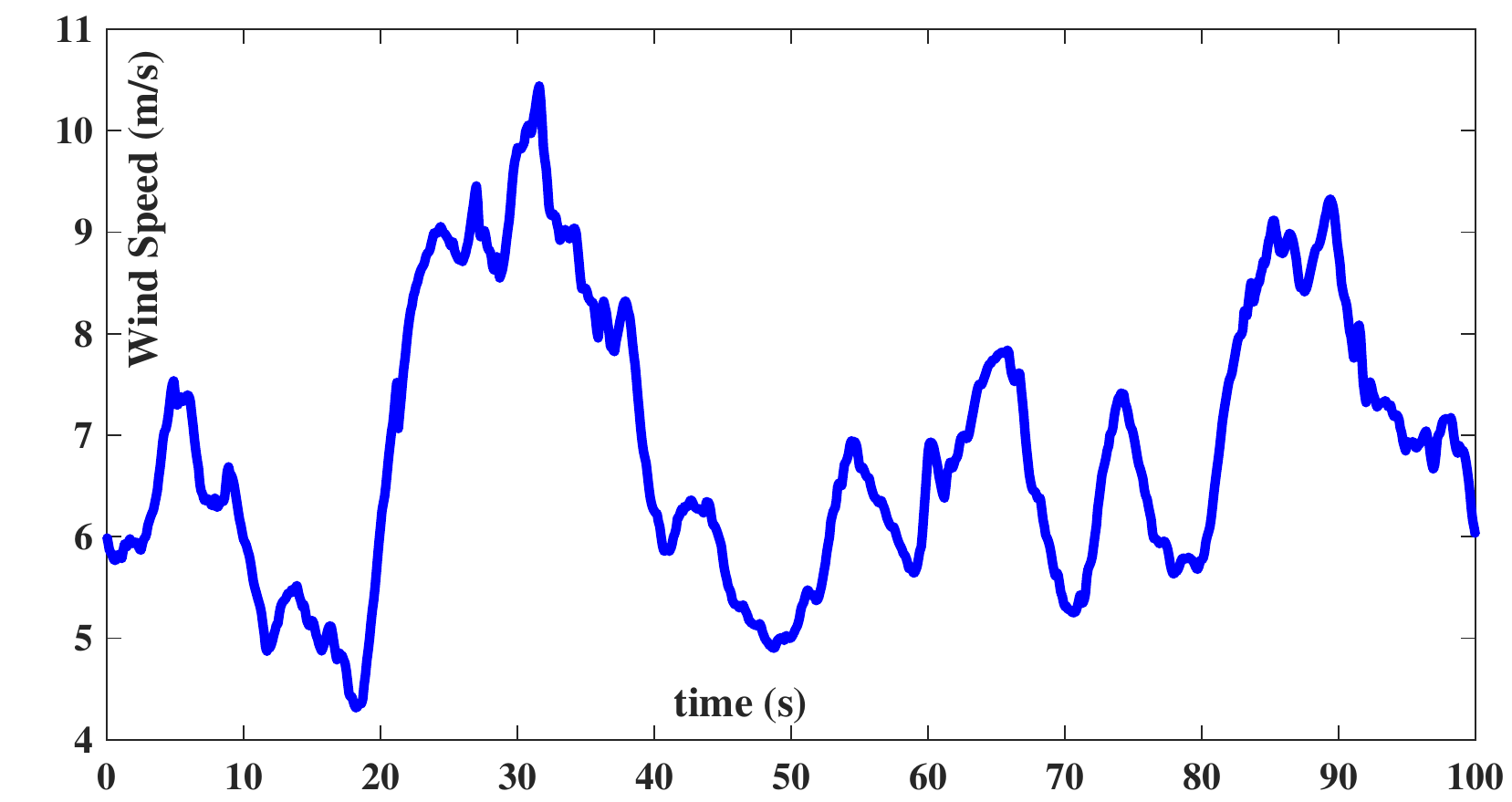}
\caption{Wind Speed.}
\label{fig_12}
\end{figure}

\begin{figure}[!t]
\centering
\includegraphics[width=3.5in]{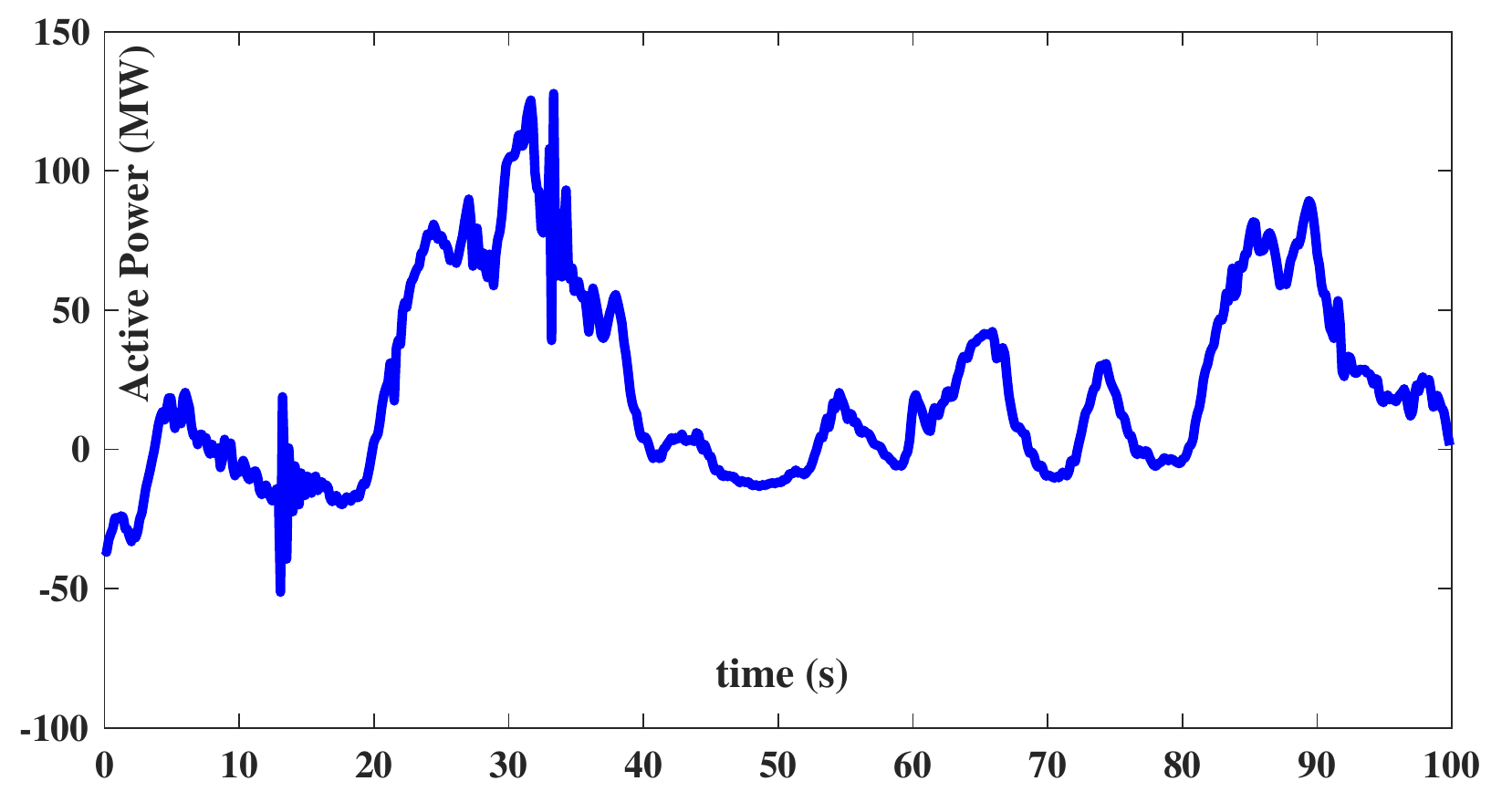}
\caption{WTG Active Power.}
\label{fig_13}
\end{figure}

\begin{figure}[!t]
\centering
\includegraphics[width=3.5in]{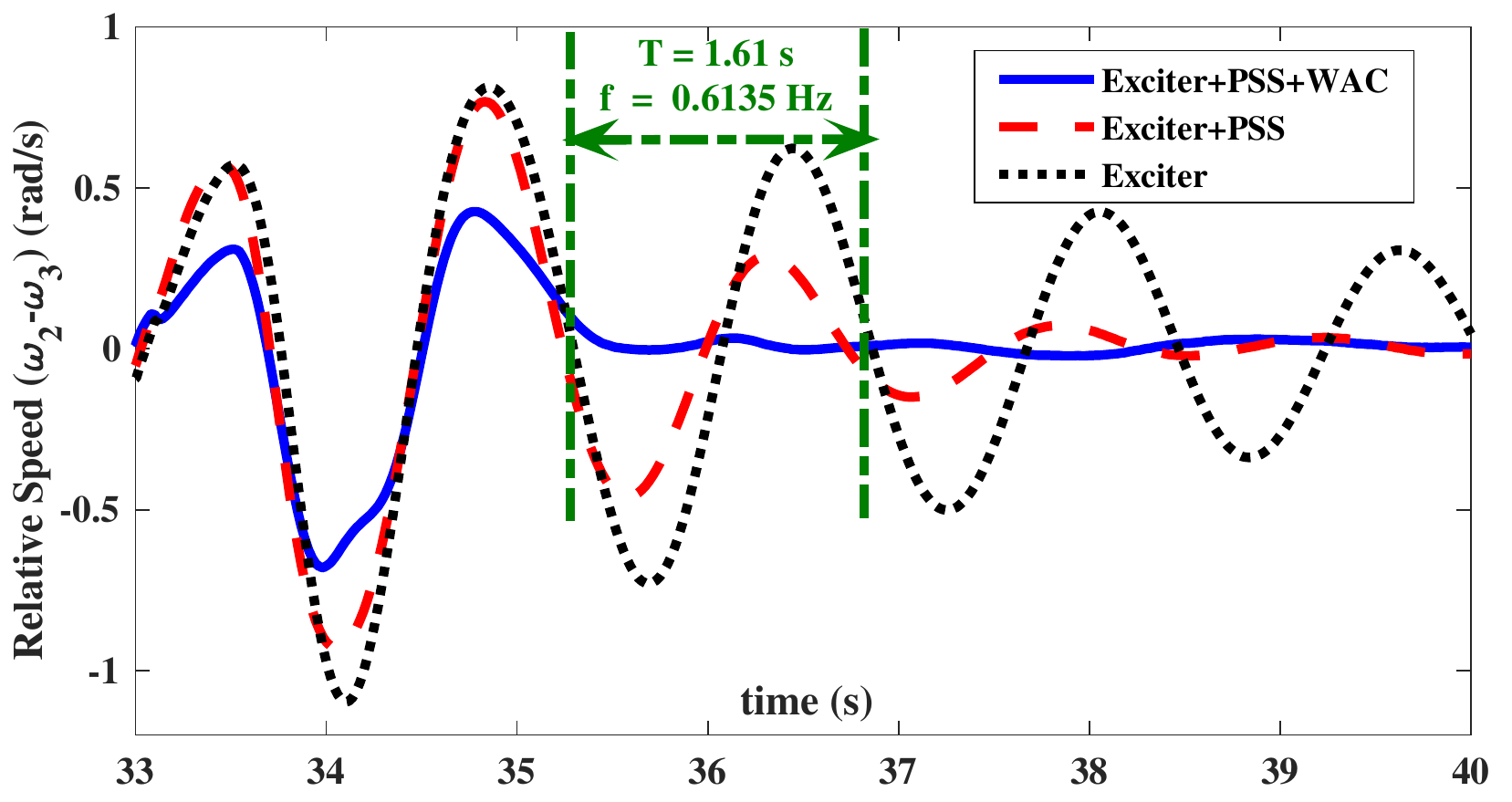}
\caption{Relative speed of generator 2 w.r.t generator 3.}
\label{fig_14}
\end{figure}

\begin{figure}[!t]
\centering
\includegraphics[width=3.5in]{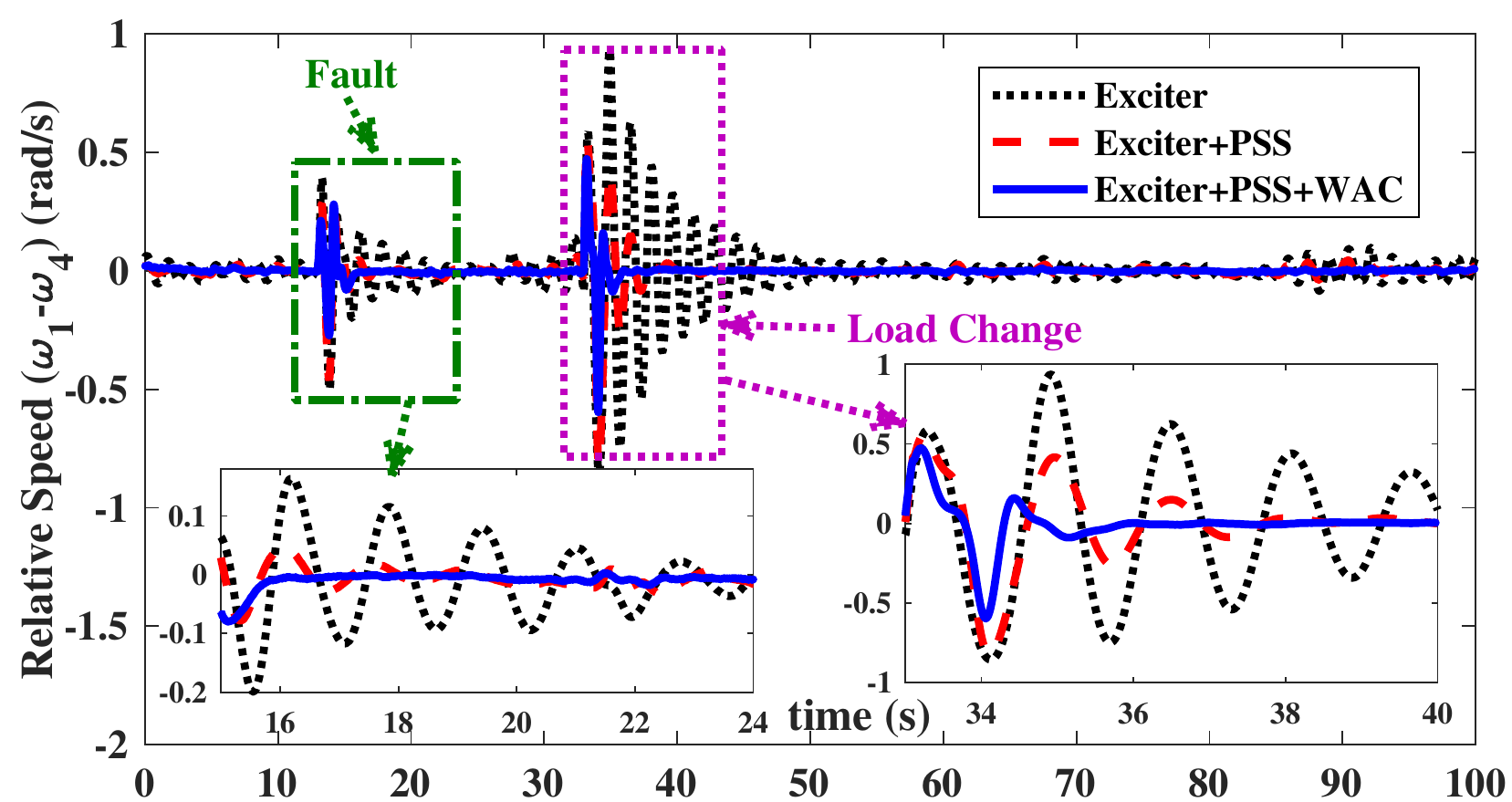}
\caption{Relative speed of generator 1 w.r.t generator 4.}
\label{fig_15}
\end{figure}

\begin{figure}[!t]
\centering
\includegraphics[width=3.5in]{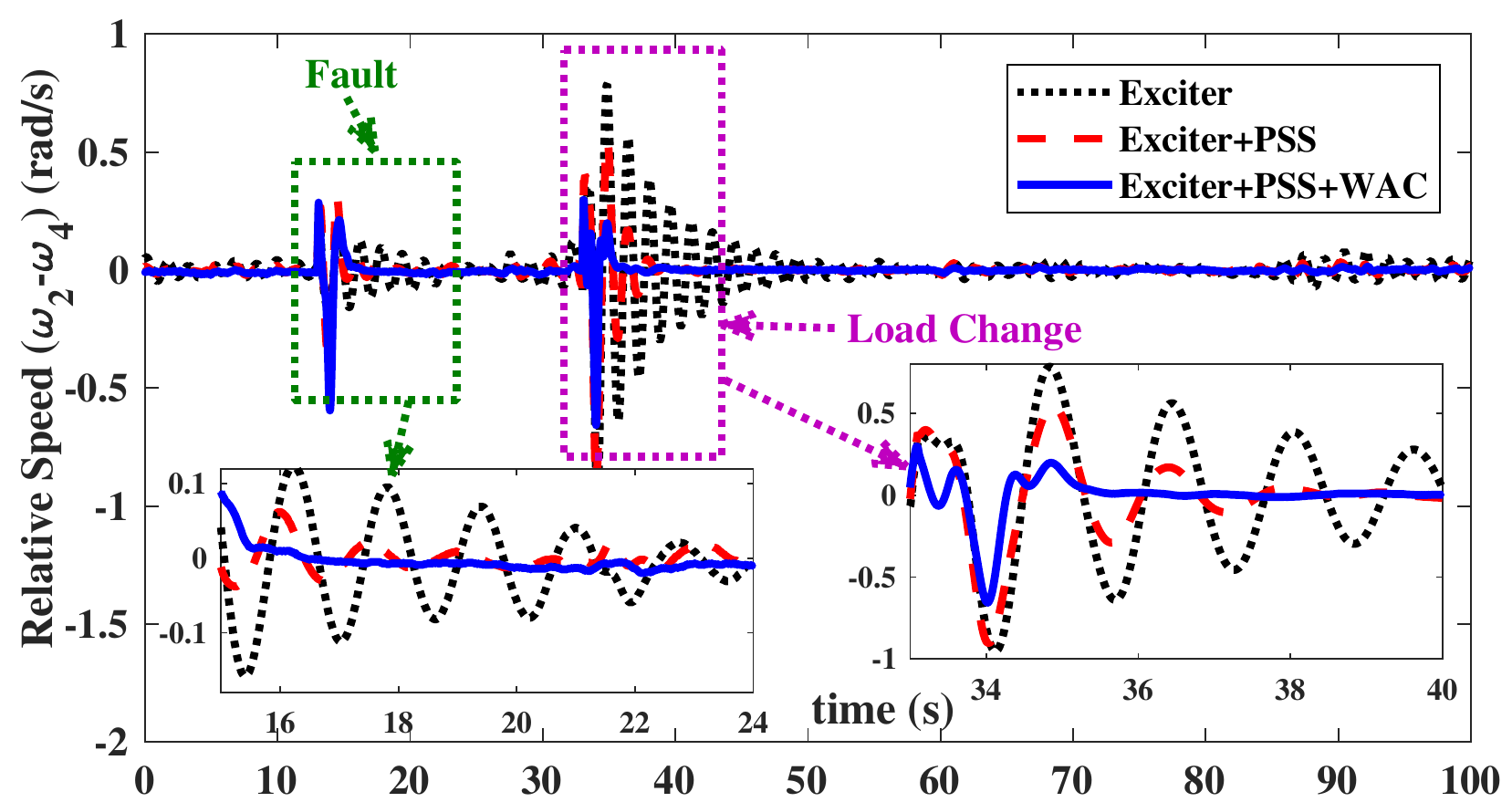}
\caption{Relative speed of generator 2 w.r.t generator 4.}
\label{fig_16}
\end{figure}

\begin{figure}[!t]
\centering
\includegraphics[width=3.5in]{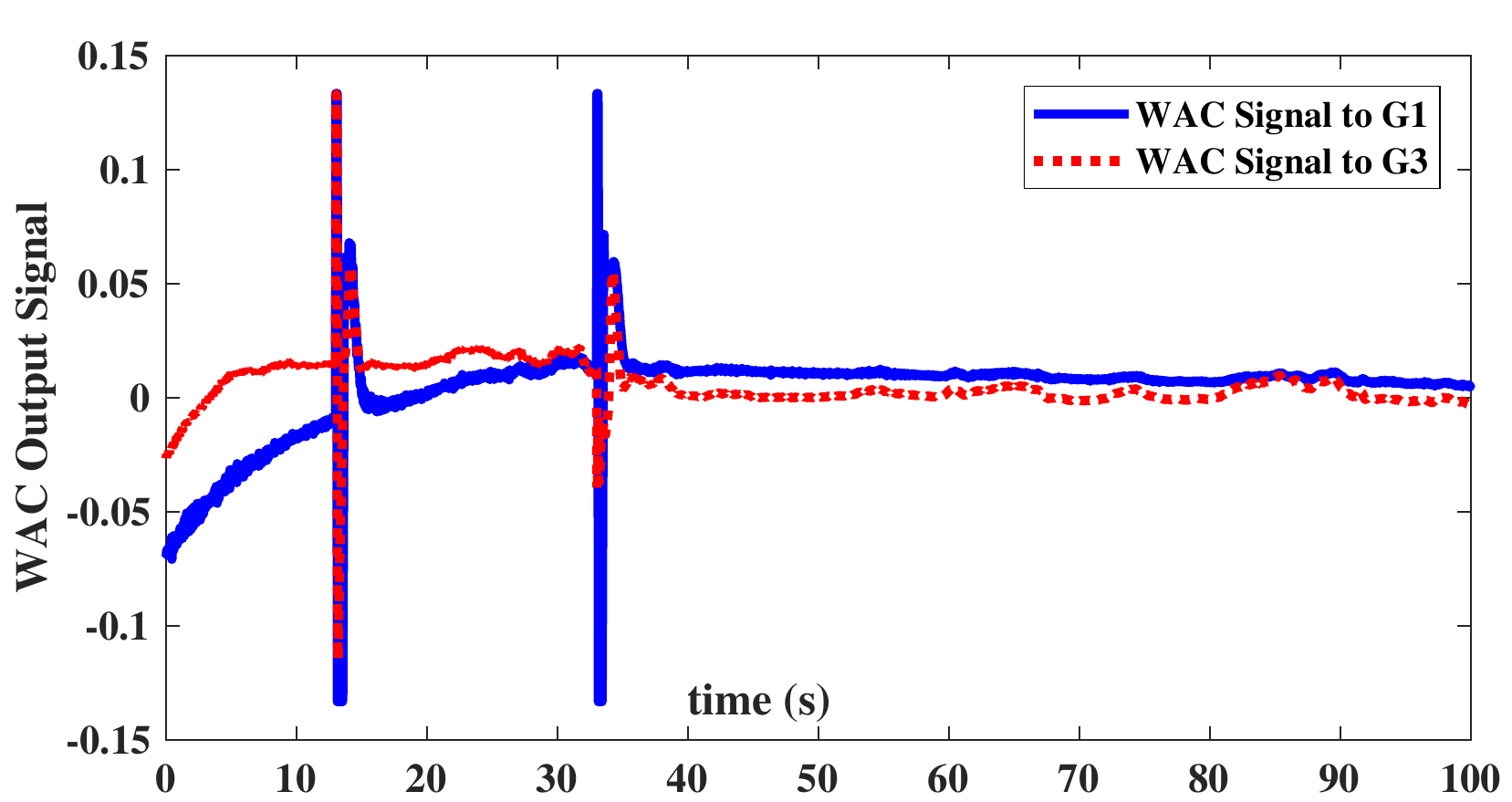}
\caption{WAC input signal to generator 1 and 3.}
\label{fig_17}
\end{figure}

Based on coherency grouping as shown in Table \ref{table_1}, it can be seen that generators 1 and 2 are in one group and generators 3 and 4 are in the other group, so two WACs are required which are to be placed one in each group. The signal selection for WAC control loop is discussed in section II. The simulation results with the proposed controller are compared with the results from a system with both Exciter and PSS, and with Exciter only. Fig. \ref{fig_12} shows the wind speed profile used for simulations. Fig. \ref{fig_13} shows the active power. It is seen that (see Fig. \ref{fig_13}) injection of the variable active power of WTG into the grid changes operating condition of the power system, which initiates synchronous generators to oscillate. At this time, a disturbance is created by initiating a fault on bus-8 at 13 sec for a duration of 0.1 sec and another disturbance is created at 33 sec by dropping load connected to bus-9 for a duration of 0.1 sec. Fig. \ref{fig_14} shows the relative speed of generator 2 w.r.t generator 3, Fig. \ref{fig_15} show the speed of generator 1 w.r.t generator 4, Fig. \ref{fig_16} shows the relative speed of generator 2 w.r.t generator 4, and Fig. \ref{fig_17} shows the WAC controlling signal to generator 1 and generator 3.

From the above results, with the WAC controlling signal (Fig. \ref{fig_17}) to generator 1 and generators 3, it is observed that the oscillations are effectively damped (Fig. \ref{fig_14} to Fig. \ref{fig_16}). From Fig. \ref{fig_14}, it is seen that the frequency of inter-area oscillation is 0.6135Hz (approx.). Since the control loop is selected for inter-area oscillation (0.6038Hz), the inter-area oscillations are damped out effectively. From these results, it can be concluded that the proposed algorithm based on the online coherency grouping performs very effectively when compared to PSS working alone, and can be implemented online.

\subsection{Implementation test results on IEEE 39-bus system}
\begin{figure}[!t]
\centering
\includegraphics[width=3.5in]{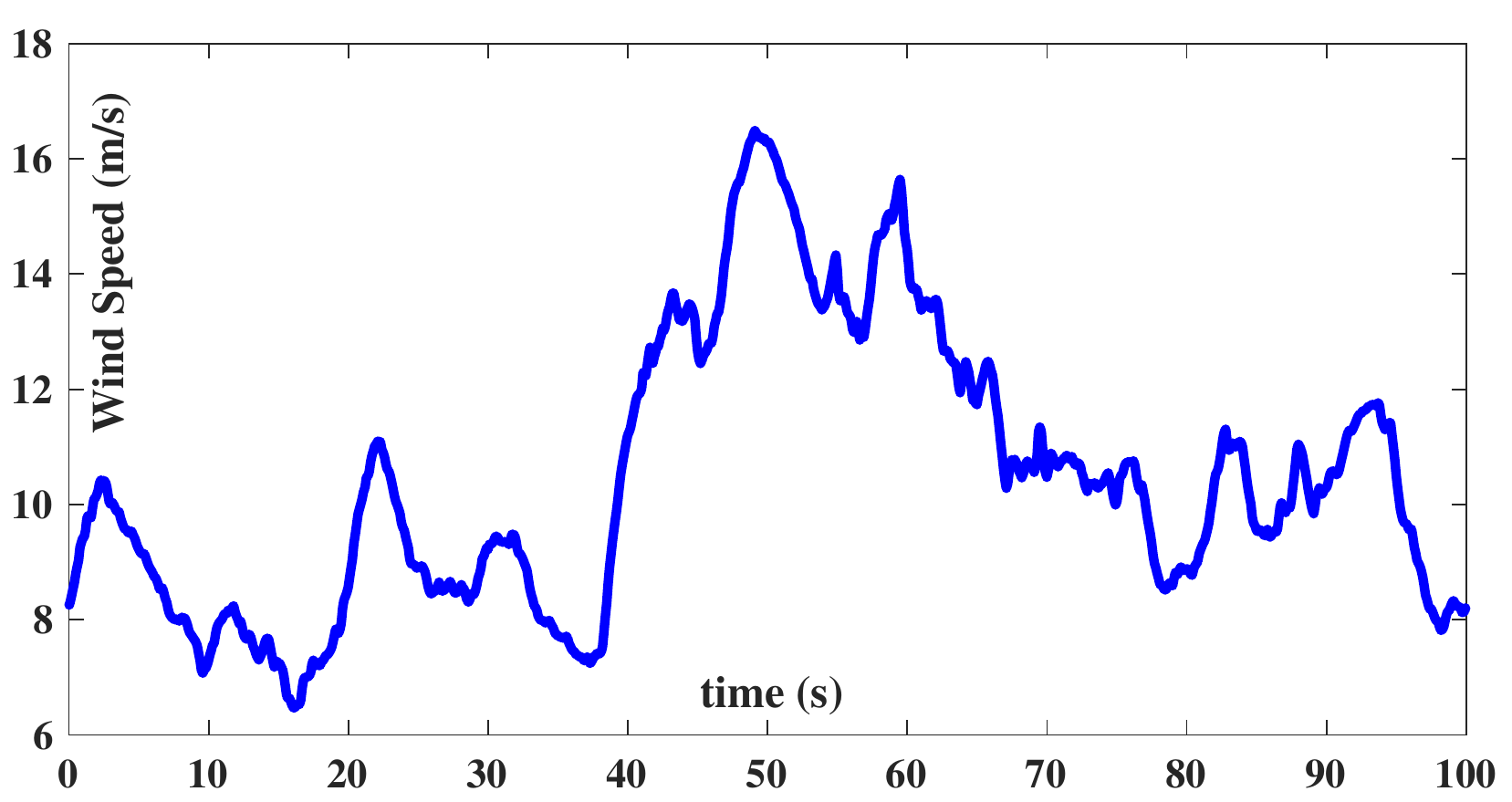}
\caption{Wind Speed.}
\label{fig_18}
\end{figure}

\begin{figure}[!t]
\centering
\includegraphics[width=3.5in]{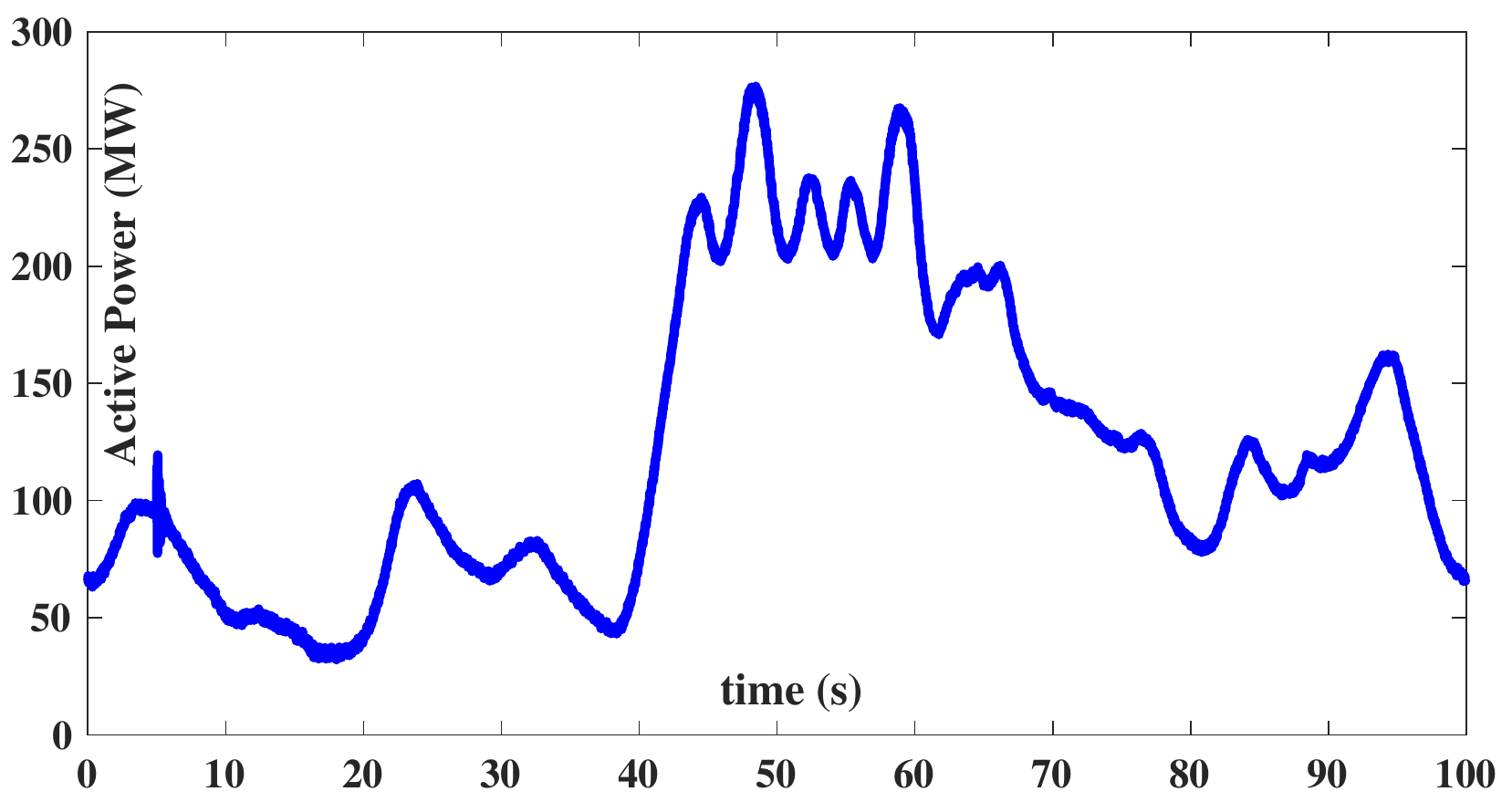}
\caption{WTG Active Power.}
\label{fig_19}
\end{figure}

\begin{figure}[!t]
\centering
\includegraphics[width=3.5in]{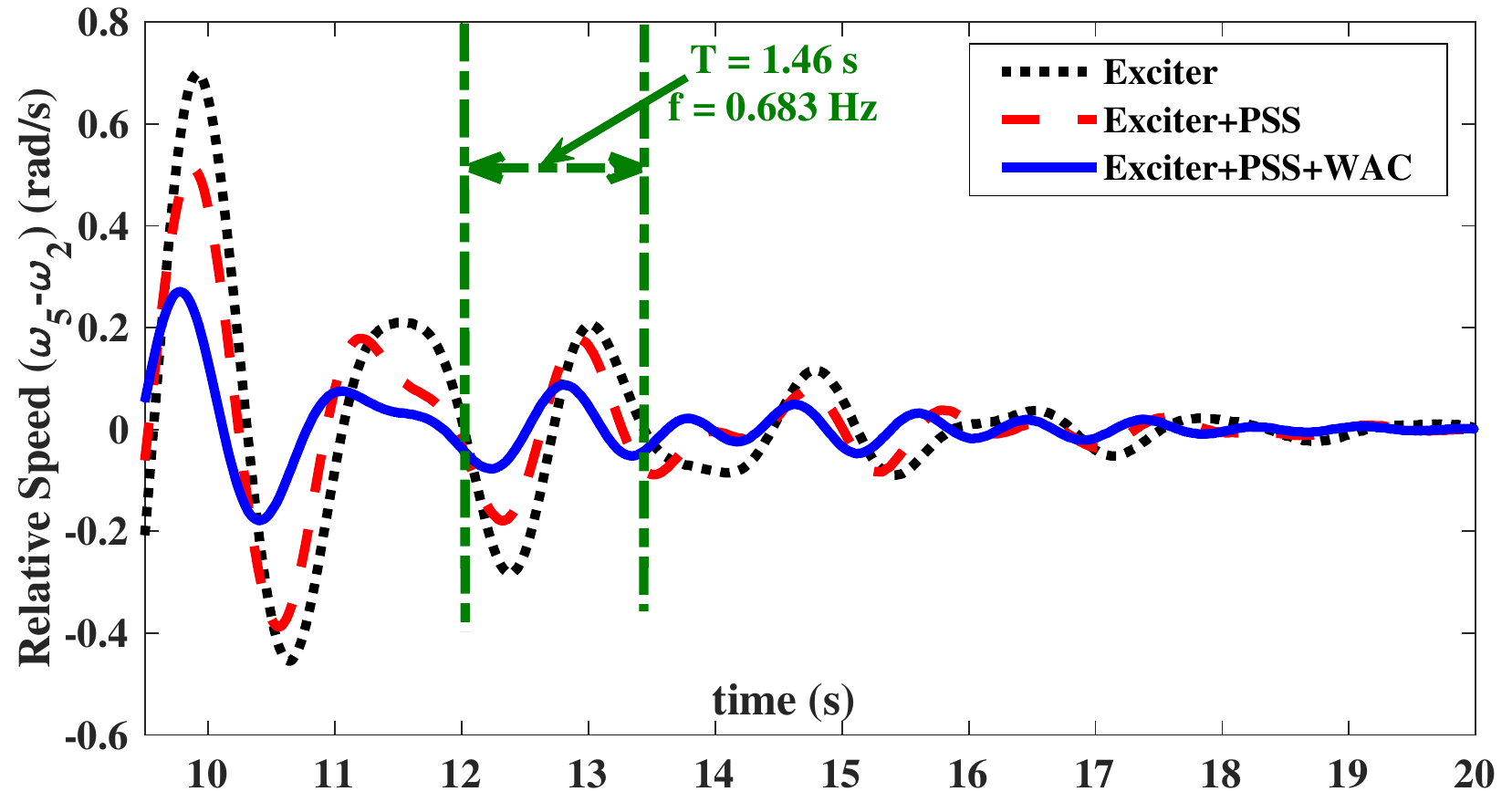}
\caption{Relative speed of generator 5 w.r.t generator 2.}
\label{fig_20}
\end{figure}

\begin{figure}[!t]
\centering
\includegraphics[width=3.5in]{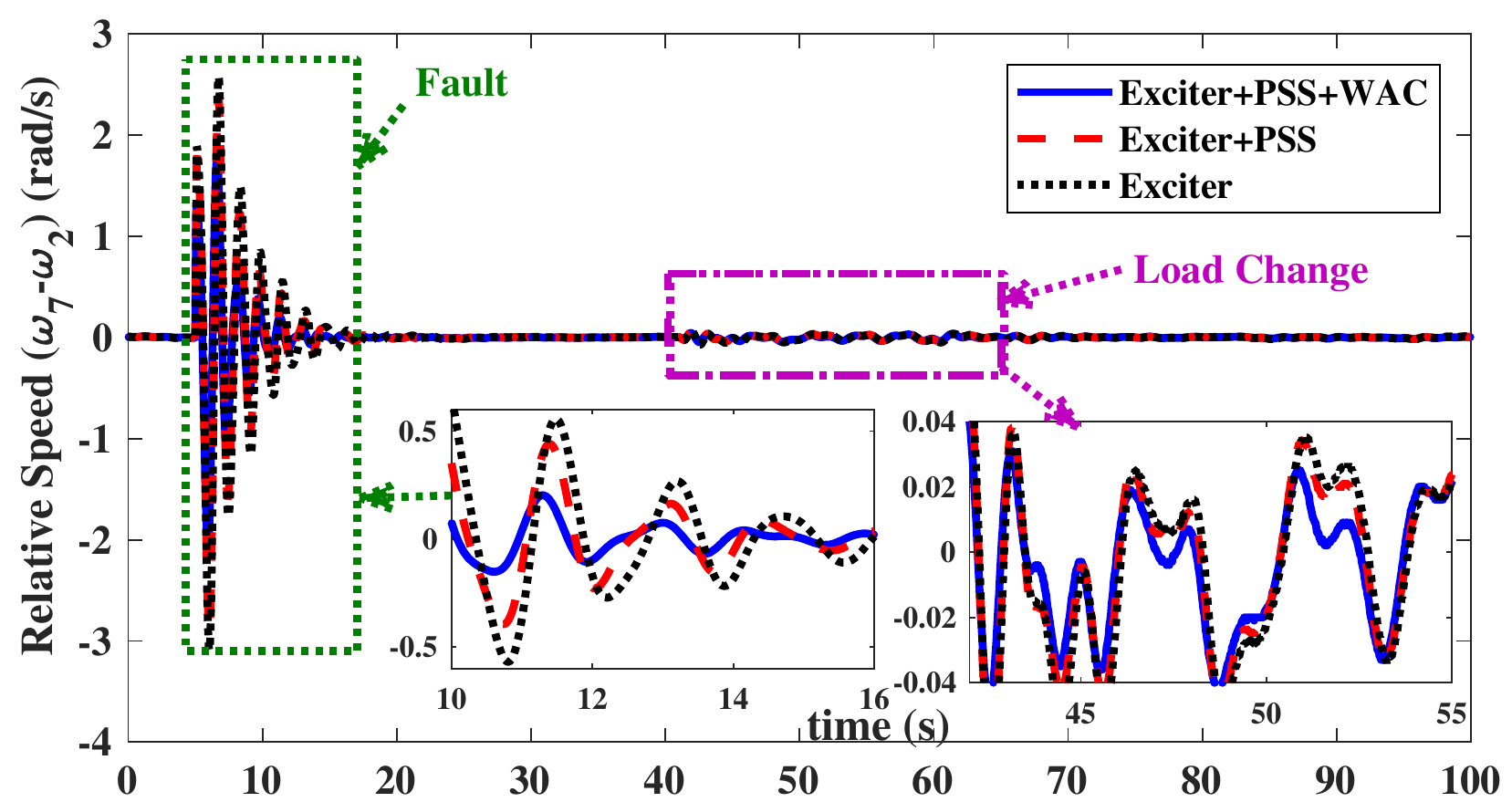}
\caption{Relative speed of generator 7 w.r.t generator 2.}
\label{fig_21}
\end{figure}

\begin{figure}[!t]
\centering
\includegraphics[width=3.5in]{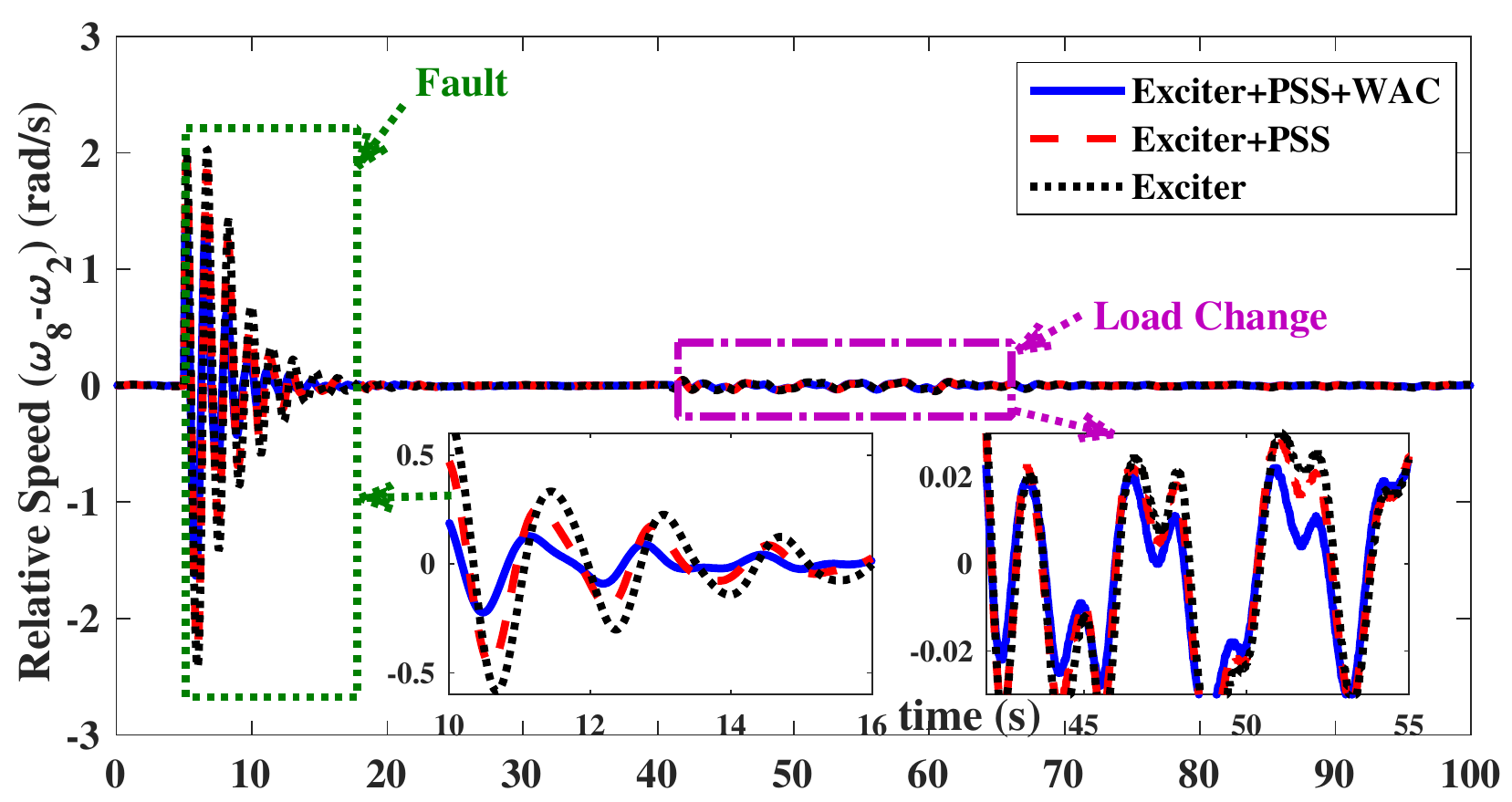}
\caption{Relative speed of generator 8 w.r.t generator 2.}
\label{fig_22}
\end{figure}

\begin{figure}[!t]
\centering
\includegraphics[width=3.5in]{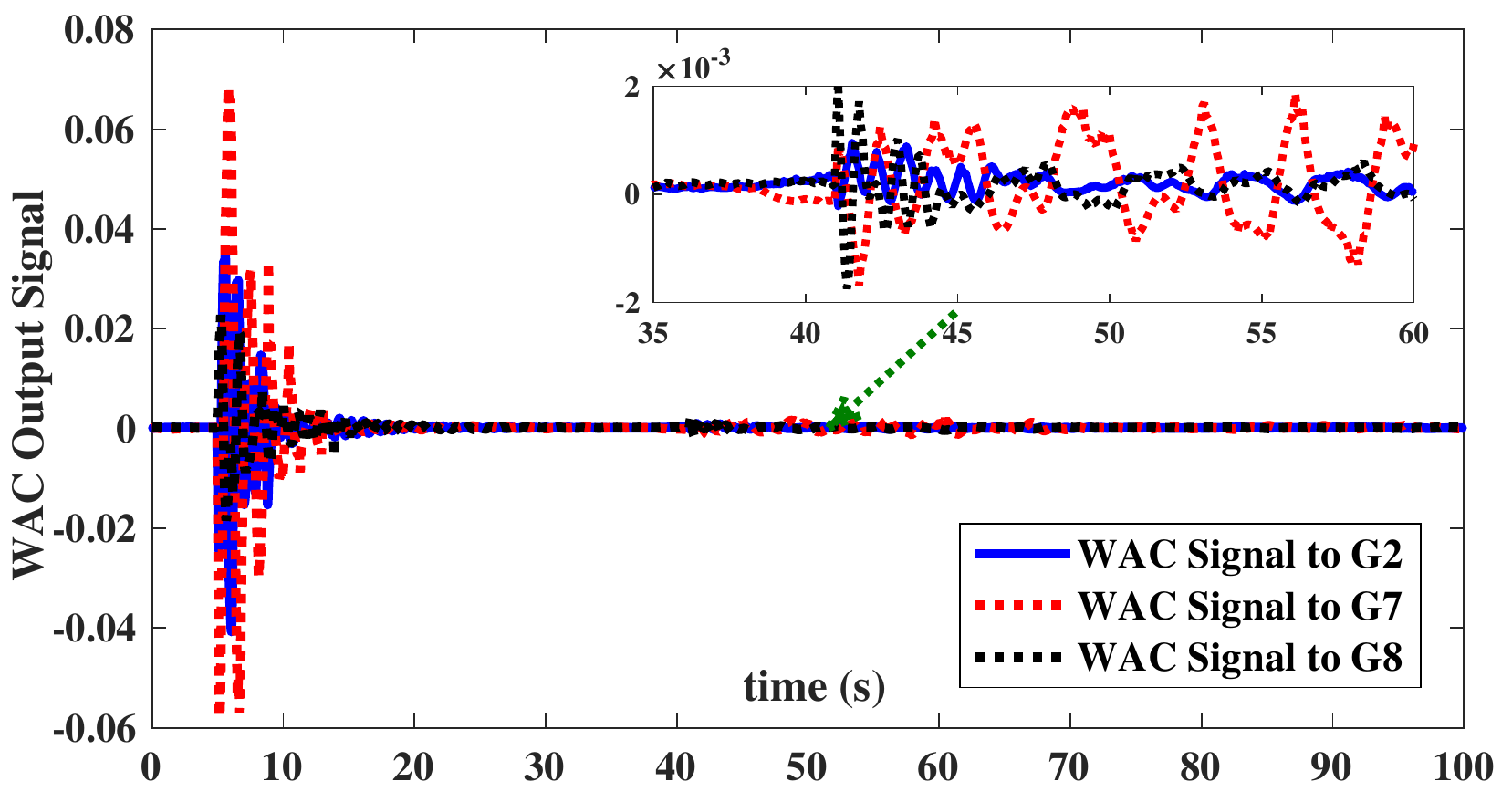}
\caption{WAC input signal to generator 2, 7, and 8}
\label{fig_23}
\end{figure}

Based on coherency grouping as shown in Table \ref{table_1}, generators 4, 5, 6, 7, and 9 are in one group (group-4), generators 1, and 8 are in group-1, generators 2, 3 are in group-3, so three WAC controllers are required which are to be placed one in each group. The signal selection for WAC control loop is shown in Table \ref{table_2a}. The simulation results with the proposed controller are compared with the results from a system with both Exciter and PSS, and with Exciter only. Fig. \ref{fig_18} shows the wind speed profile used for simulations. Fig. \ref{fig_19} shows the active power. It is seen that (see Fig. \ref{fig_19}), injection of the variable active power of WTG into the grid changes the operating condition of the power system initiating oscillations in synchronous generators. At this time, a disturbance is created by initiating a fault on bus-14 at 5 sec for a duration of 0.1 sec and another disturbance is created at 41 sec by dropping load connected to bus 26 for a duration of 0.1 sec. Fig. \ref{fig_20} shows the relative speed of generator 5 w.r.t generator 2, Fig. \ref{fig_21} show the speed of generator 7 w.r.t generator 2, Fig. \ref{fig_22} shows the relative speed of generator 8 w.r.t generator 2, and Fig. \ref{fig_23} shows the WAC controlling signal to generator 2, 7, and 8. From Fig. \ref{fig_20} to Fig. \ref{fig_22}, it can be seen that after 17sec the oscillations are damped effectively hence the WAC output is dropped.
\begin{table}[!t]
\renewcommand{\arraystretch}{1.3}
\centering
\caption{IEEE 39-BUS CONTROL LOOP \\ (FAULT ON BUS-14)}
\label{table_2a}
\begin{tabular}{*9c}
\hline
\toprule
 & Control Loop & Residue \\
\hline
Group-1 & $\Delta\omega_{10}$ $\rightarrow$ $u_7$ & 0.678 \\
Group-2 & $\Delta\omega_7$ $\rightarrow$ $u_8$ & 0.49 \\
Group-3 & $\Delta\omega_6$ $\rightarrow$ $u_2$ & 1 \\
Group-4 & $\Delta\omega_1$ $\rightarrow$ $u_{10}$ & 0.027 \\
\hline
\multicolumn{3}{c}{Computational Time (Signal Selection) = 3.16s} &  \\
\multicolumn{3}{c}{Computational Time (WAC Output Signal) = 0.0192s} &  \\
\hline
\end{tabular}
\end{table}

From the above results, with the WAC controlling signal (Fig. \ref{fig_23}) to generator 2, 7, and 8, it is observed that the oscillations are effectively damped (Fig. \ref{fig_20} to Fig. \ref{fig_22}). From Fig. \ref{fig_20} it is seen that the frequency of inter-area oscillation is 0.683Hz (approx.). Since the control loop is selected for inter-area oscillation (0.6038Hz), the inter-area oscillations are damped out effectively. Hence it can be concluded that the proposed algorithm based on the online coherency grouping performs very effectively when compared to PSS working alone, and can be implemented online.

\section{Conclusion}
In this paper an online coherency based WAC is proposed. The proposed WAC algorithm is an efficient way of damping inter-area modes. The location of WAC is decided based on online coherency grouping and signal selection is based on online residue analysis. The spectral clustering and residue using MIMO system identification are computed during the simulation, so any changes in system operating conditions especially due to renewable energy sources are taken into consideration. The WAC gains calculation and state estimation are performed online, hence the controller parameters are updated based on system conditions. The efficacy of the proposed method is verified by testing it on WTG integrated two-area and IEEE 39-bus power system, models. The proof of concept illustrates that with the proposed WAC, inter-area modes can be damped much effectively.

\appendices
\section{Parameters of Wind Turbine Generator \cite{ref15}}
\centering
\begin{tabular}{|c|c|}
\hline
\toprule
\textbf{Parameter Name} &  \textbf{Value}\\
\hline
Rated generator power & 2.2 MVA\\
\hline
Rated turbine power & 2.0 MW\\
\hline
Generator speed at rated  & 1.2 p.u.\\
turbine speed (p.u.) & \\
\hline
Rated wind speed & 12.0 m/s\\
\hline
Cut-in wind speed & 6.0 m/s\\
\hline
\end{tabular}

\section{Parameters of Wound Rotor Induction Machine}
\begin{tabular}{|c|c|}
\hline
\toprule
\textbf{Parameter Name} &  \textbf{Value}\\
\hline
Rated stator voltage (L-L RMS) & 0.69 kV\\ 
\hline
Turn ratio (rotor over stator) & 2.6377\\
\hline
Rated MVA & 2.2 MVA\\
\hline
Stator resistance & 0.00462 p.u.\\ 
\hline
Stator leakage reactance & 0.102 p.u.\\ 
\hline
Unsaturated magnetizing reactance & 4.348 p.u.\\  
\hline
First cage rotor resistance & 0.006 p.u.\\
\hline
First cage rotor leakage reactance & 0.08596 p.u.\\ 
\hline
Inertia constant & 1.5 MWs/MVA\\
\hline
\end{tabular}
\bibliographystyle{IEEEtran}
\bibliography{Ref.bbl}

\vspace{-12mm}
\justify
\begin{IEEEbiography}[{\includegraphics[width=1in,height=1.25in,clip,keepaspectratio]{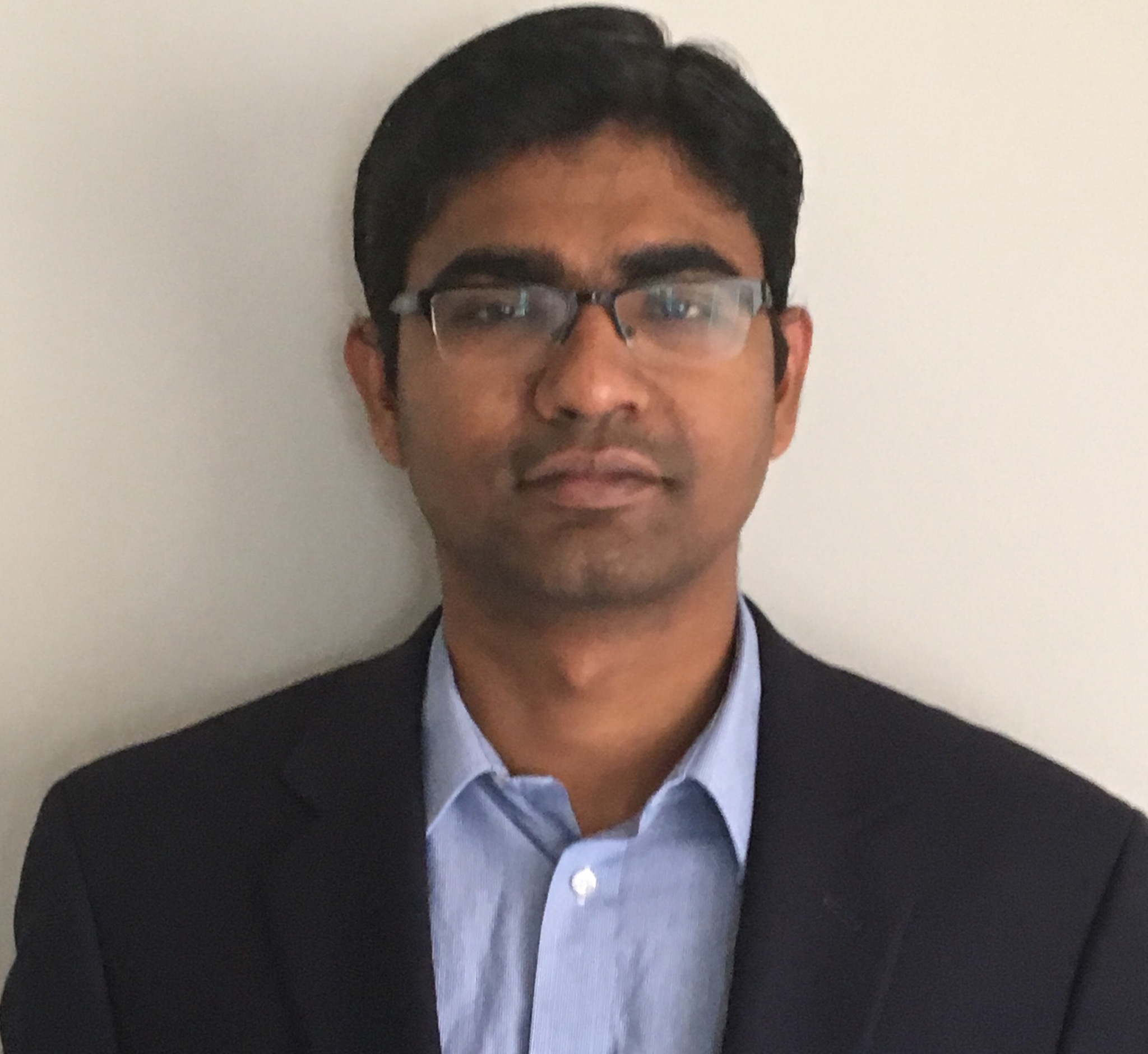}}]{A. Thakallapelli}(S'14) received his B.Tech degree in Electrical Engineering from Acharya Nagarjuna University in 2010 and the M.Tech degree in Electrical Engineering from the Veermata Jijabai Technological Institute in 2012. He is currently working toward the Ph.D degree in Electrical Engineering from the Department of Electrical and Computer Engineering, University of North Carolina at Charlotte. His research interests include wide-area control, reduced order modeling, power system stability and renewable energy.
\end{IEEEbiography}
\vspace{-13 mm}
\begin{IEEEbiography}[{\includegraphics[width=1in,height=1.25in,clip,keepaspectratio]{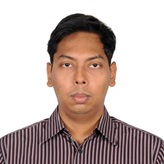}}]{S. J. Hossain} (S’ 16) received the B.Sc in Electrical and Electronic Engineering from Bangladesh University of Engineering and Technology, Bangladesh, in 2013. He worked as a software engineer in Samsung Research and Development institute Bangladesh from 2013 to 2015. He is currently a PhD candidate at University of North Carolina at Charlotte. His research interests are distributed energy systems integration, modeling and control, and wide area monitoring,
optimization and control of power system.
\end{IEEEbiography}
\vspace{-13 mm}
\begin{IEEEbiography}
[{\includegraphics[width=1in,height=1.25in,clip,keepaspectratio]{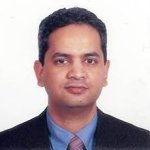}}]{S. Kamalasadan}(S'01, M'05, SM'17) received his Ph.D. in Electrical Engineering from the University of Toledo, Ohio in 2004, M.Eng in Electrical Power Systems Management, from the Asian Institute of Technology, Bangkok Thailand in 1999 and B Tech. degree in Electrical and Electronics from the University of Calicut, India in 1991. He is currently working as a Professor in the department of electrical and computer engineering at the University of North Carolina at Charlotte. He has won several awards including the NSF CAREER award and IEEE best paper award. His research interests include Intelligent and Autonomous Control, Power Systems dynamics, Stability and Control, Smart Grid, Micro-Grid and Real-time Optimization and Control of Power System. 
\end{IEEEbiography}
\end{document}